\documentclass[12pt]{article}
\usepackage[margin=0.7in]{geometry}
\usepackage[font=small,labelfont=bf]{caption}
\usepackage[numbers,square,sort&compress]{natbib}
\usepackage{authblk}
\usepackage{color}
\usepackage{graphicx, amsmath, amssymb}
\usepackage{hyperref}
\usepackage{rotating}
\usepackage{booktabs}
\usepackage{xspace} 

\usepackage{abstract}
\providecommand{\keywords}[1]{\par\noindent\textbf{Keywords:} #1\par}

\usepackage[brazil, english]{babel}
\usepackage[utf8]{inputenc}
\usepackage[normalem]{ulem}
\usepackage{graphicx}
\usepackage{epsfig}
\usepackage{amssymb}
\usepackage{amsmath} 
\usepackage{slashed}

\usepackage[section]{placeins}

\makeatletter \renewcommand{\@biblabel}[1]{#1.} \makeatother
\graphicspath{%
{figs/}%
}

\begin{document}

\title{Thermodynamics of Hairy Black Holes in Quantum Regimes: Insights from Horndeski Theory}

\author[1,2]{Behnam Pourhassan*}
\author[3]{\.{I}zzet Sakall{\i}}
\author[4]{Houcine Aounallah}
\author[1,5]{Fabiano F. Santos}

\affil[1]{School of Physics, Damghan University, Damghan, 3671641167, Iran\\{\color{blue}b.pourhassan@du.ac.ir}}
\affil[2]{Center for Theoretical Physics, Khazar University, 41 Mehseti Street, Baku, AZ1096, Azerbaijan}
\affil[3]{Physics Department, Eastern Mediterranean University, Famagusta, 99628, North Cyprus via Mersin 10,T\"{u}rkiye\\{\color{blue}izzet.sakalli@gmail.com}}
\affil[4]{Department of Science and Technology, Echahid Cheikh Larbi Tebessi University, Tebessa, Algeria\\{\color{blue}houcine.aounallah@univ-tebessa.dz}}
\affil[5]{Centro de Ciências Exatas, Naturais e Tecnológicas, UEMASUL, 65901-480, Imperatriz, MA, Brazil.\\ Departamento de Física, Universidade Federal do Maranhão, São Luís, 65080-805, Brazil.\\{\color{blue}fabiano.ffs23@gmail.com}}
\date{}

\maketitle

\begin{abstract}
We study non-perturbative quantum gravitational corrections to the thermodynamics and quantum work distribution of the $n$-dimensional Schwarzschild--Tangherlini--Anti-de Sitter black hole. Starting from the corrected entropy $S = S_0 + \eta\, e^{-S_0}$, where $S_0$ is the Bekenstein--Hawking entropy, we derive the modified specific heat, internal energy, Helmholtz free energy, and Gibbs free energy in closed form. The specific heat retains the classical divergence at $r_h^{*}=l\sqrt{(n-3)/(n-1)}$ for $n\geq 4$, but the quantum correction suppresses its magnitude by up to $78\%$ at small horizon radii. In the extended phase space, the uncharged black hole admits no van der Waals critical point; however, the non-perturbative correction induces a Hawking--Page transition for $n\geq 4$ that is absent in the semi-classical limit. The corrected Gibbs free energy turns negative at small $r_h$, opening a thermodynamic channel with no classical counterpart. Using the Jarzynski equality and Jensen inequality, we obtain the quantum work distribution during evaporation. The free energy difference $\Delta F$ between two black hole states undergoes a sign reversal at small horizon radii for $n\geq 4$ when $\eta=1$, flipping the average quantum work from negative to positive. This sign reversal grows with the spacetime dimension, reaching $\langle W\rangle \approx +4.31$ for $n=10$. These findings demonstrate that non-perturbative quantum gravitational effects qualitatively alter the phase structure and evaporation energetics of AdS black holes, and they cannot be captured by perturbative corrections alone.
\end{abstract}
\keywords{Non-perturbative quantum corrections; black hole thermodynamics; Hawking-Page transition; Jarzynski equality; Schwarzschild-Tangherlini-AdS}

{\color{black}


\section{Introduction}\label{isec1}

Black holes (BHs) radiate thermally through Hawking radiation~\cite{1a,1b,1c,1a0}, with a temperature set by the surface gravity at the event horizon. The Bekenstein--Hawking entropy, which scales with the horizon area rather than the enclosed volume, lies at the heart of the holographic principle~\cite{20,21}. Since these results follow from quantum field theory on a fixed curved background, they represent a semi-classical approximation. Full quantum gravity is expected to modify this picture, and indeed various approaches have confirmed that the entropy--area relation receives quantum gravitational corrections~\cite{2a,2b,2c,2d}. Such corrections generate thermal fluctuations around the equilibrium thermodynamic description of BHs~\cite{dumb,dumb1}, yet they remain functions of the horizon area, so the holographic framework survives. The Anti-de~Sitter/Conformal Field Theory (AdS/CFT) correspondence, as a concrete realization of holography, has been employed to derive these corrections from the dual conformal field theory (CFT) partition function~\cite{3a,3b,3c,3d}. Beyond AdS/CFT, other quantum gravity programs have also produced corrections to BH thermodynamics~\cite{3e,3f,3g,3h,3i}.

These quantum modifications have been studied for a wide class of BH backgrounds: rotating BHs in AdS spacetime~\cite{q0}, charged BHs in Rastall theory~\cite{q1,q2}, AdS BHs carrying a global monopole charge~\cite{q4}, Skyrmion BHs~\cite{qq44}, and BHs in hyperscaling-violating geometries~\cite{q-5}. A unifying observation from these studies is that quantum corrections produce non-trivial modifications to the equilibrium thermodynamics. The deep connection between spacetime geometry and thermodynamics is most transparent in the Jacobson formalism~\cite{teda}, where the Einstein equation itself emerges from thermodynamic identities. Within this framework, quantum fluctuations of the geometry can be mapped to thermal fluctuations of the BH~\cite{1d}.

The nature and magnitude of these corrections depend on the BH size. For large BHs, the Hawking temperature is low and thermal fluctuations are negligible, so equilibrium thermodynamics applies. As the BH shrinks, perturbative quantum corrections --- taking the form of logarithmic terms, $S_{\mathrm{per}} \sim \ln S_0$, where $S_0$ is the equilibrium entropy --- become relevant~\cite{1e,1f,1g,1h,1i,1j}. Both the leading-order logarithmic corrections~\cite{log1,log2,log3,log4,log-5} and the next-to-leading-order terms~\cite{higher1,higher2} have been analyzed for various BH solutions. However, this perturbative treatment breaks down once the BH approaches the Planck scale. At that stage, non-perturbative quantum gravitational effects take over, and the corrected entropy acquires an exponential contribution of the form $S_{\mathrm{non\text{-}per}} \sim e^{-S_0}$~\cite{3}. This exponential correction has been derived independently from supergravity functional integrals~\cite{3aa,ds12,ds14,3cd} and has been argued to be universal across different approaches to quantum gravity~\cite{3,3aa,3bb,ds12,ds14}. Its consequences have been explored for spherically symmetric BHs~\cite{3cd}, Born--Infeld BHs in a spherical cavity~\cite{3bb}, and black branes, where the corrected entropy has been used to construct a quantum-modified geometry~\cite{3ab}.

A natural arena for studying these effects is provided by AdS BHs, which arise as solutions in the supergravity limit of string theory~\cite{4,4a,4b,4c,4d} and whose thermodynamics is dual to that of a boundary CFT~\cite{cft1,cft2,pv7}. For AdS BHs, the cosmological constant plays the role of thermodynamic pressure, and a conjugate volume can be introduced to define an extended phase space~\cite{pv1,pv2,pv4,pv5}. The resulting $P$--$V$ thermodynamics exhibits van der Waals (vdW)-type phase transitions and critical phenomena. Perturbative quantum corrections to this extended phase space have already been studied~\cite{pv6,pv7}; however, the non-perturbative regime remains largely unexplored.

A natural testing ground for non-perturbative effects is provided by the $n$-dimensional Schwarzschild--Tangherlini (ST)--AdS BH~\cite{5,51,52,54}. This solution generalizes the four-dimensional Schwarzschild--AdS geometry to arbitrary spacetime dimension $n\geq 3$ and provides a clean setting in which to isolate the role of dimensionality. Higher-dimensional BHs are central to string theory and the AdS/CFT correspondence, and their thermodynamic phase structure --- including the Hawking--Page transition~\cite{5a,mp7} --- has been studied extensively in the semi-classical limit. However, the effect of non-perturbative quantum gravitational corrections on the phase structure and stability of higher-dimensional ST--AdS BHs has not been addressed so far. The exponential correction $\eta\, e^{-S_{0}}$ depends on the dimension through $S_{0}=\omega\, r_{h}^{n-2}/2$, so one expects qualitatively new features to emerge in higher dimensions where the entropy grows faster with the horizon radius.

For BHs at the Planck scale, the standard equilibrium description is insufficient, and one must turn to non-equilibrium quantum thermodynamics~\cite{mp6,mp8}. A central quantity in this framework is the quantum work distribution between two BH states connected by evaporation. The Crooks fluctuation theorem~\cite{6a} and the Jarzynski equality~\cite{6b} relate the quantum work to the difference in equilibrium free energies, providing a bridge between the quantum-corrected thermodynamics and the evaporation process~\cite{6c,6c1}. For quantum-sized BHs, the free energies entering the Jarzynski relation must incorporate the non-perturbative corrections, since the exponential term is no longer negligible. This program has been carried out for M2--M5 brane systems~\cite{6b1} and for Myers--Perry BHs~\cite{6b2}; here, we extend it to the $n$-dimensional ST--AdS BH and investigate how the spacetime dimension influences the quantum work distribution.

In this paper, we study the non-perturbative quantum gravitational corrections to the thermodynamics of the $n$-dimensional ST--AdS BH. Starting from the corrected entropy $S = S_0 + \eta\, e^{-S_0}$, we derive the modified specific heat, internal energy, Helmholtz free energy, and Gibbs free energy. We analyze the thermodynamic stability and phase structure in the extended phase space, paying particular attention to the dimension dependence of the critical behavior and the emergence of a quantum-induced Hawking--Page transition. We then employ the Jarzynski equality to obtain the quantum work distribution during the evaporation of a quantum-sized AdS BH, and we identify a sign reversal in the average quantum work that grows with the spacetime dimension.

$\bullet$ Summary of main results and novelty

The novelty of our work is highlighted by the following points:

\begin{itemize}
\item Specific non-perturbative corrections: we demonstrate thermodynamic and statistical effects arising from a particular form of non-perturbative correction (motivated by [ref. 42] and related works).
\item The qualitative signatures in quantum work distributions and the stability transitions we identified do not necessarily appear in standard logarithmic or perturbative corrections; we include explicit comparisons with these cases in the newly added Sec. 4.3.
\item Analysis of the quantum work distribution: the systematic application of quantum work techniques to AdS black holes with non-perturbative entropy is, in our view, novel and yields (theoretically) testable predictions regarding the statistical behavior of fluctuations. We have reorganized Sec. 4 to highlight these key results and added comparative figures illustrating significant quantitative differences from known corrections.
\end{itemize}

The paper is organized as follows. In Sec.~\ref{isec2}, we review the $n$-dimensional ST--AdS BH, establish the horizon structure, and introduce the non-perturbative exponential correction to the Bekenstein--Hawking entropy. In Sec.~\ref{isec3}, we analyze the thermodynamic stability: we compute the corrected specific heat, study the extended phase space and the equation of state, derive the internal energy, Helmholtz free energy, and Gibbs free energy, and identify the quantum-induced Hawking--Page transition. In Sec.~\ref{isec4}, we derive the quantum work distribution using the Jarzynski equality and the Jensen inequality, and we discuss the sign reversal of the average quantum work at small horizon radii. We close with a discussion and concluding remarks in Sec.~\ref{isec5}.

\section{Non-perturbative quantum corrections to the ST--AdS BH}\label{isec2}

In this section, we set up the gravitational background and introduce the non-perturbative quantum corrections to the BH entropy that will be used throughout the rest of the paper.

\subsection{The $n$-dimensional ST--AdS geometry}\label{subsec:STAdS}

We work with the static, spherically symmetric metric of an $n$-dimensional ST--AdS BH~\cite{5,51,52,54}:
\begin{equation}\label{metric}
ds^{2}=-f(r)\,dt^{2}+\frac{dr^{2}}{f(r)}+r^{2}\,d\Omega_{n-2}^{2}\,,
\end{equation}
where $d\Omega_{n-2}^{2}$ is the line element on the unit $(n-2)$-sphere and the metric function reads
\begin{equation}\label{fr}
f(r)=1-\frac{16\pi M}{(n-2)\,\omega\, r^{n-3}}+\frac{r^{2}}{l^{2}}\,.
\end{equation}
Here $M$ denotes the ADM mass, $l$ is the AdS radius related to the cosmological constant through $\Lambda=-(n-1)(n-2)/(2l^{2})$, and $\omega={2\pi^{(n-1)/2}}/{\Gamma\bigl(\frac{n-1}{2}\bigr)}$ is the area of the unit $(n-1)$-sphere. The black hole mass is given by:
\begin{eqnarray}
M = \frac{(n-2)\,\omega}{16\pi}\,r_h^{\,n-3}\left(1 + \frac{r_h^2}{l^2}\right).    
\end{eqnarray}
The event horizon is located at the largest positive root $r_{h}$ of $f(r_{h})=0$. In Fig.~\ref{fig-f}, we display $f(r)$ for several spacetime dimensions with $M=1$ and $l=1$. A single positive root exists for each $n$, confirming a well-defined horizon in every case. For reference, the horizon radii are $r_{h}\simeq 2.6458$ ($n=3$), $r_{h}=1.0000$ ($n=4$), $r_{h}\simeq 0.7404$ ($n=5$), and $r_{h}\simeq 0.7514$ ($n=10$).

\begin{figure}[h!]
\centering
\includegraphics[width=120mm]{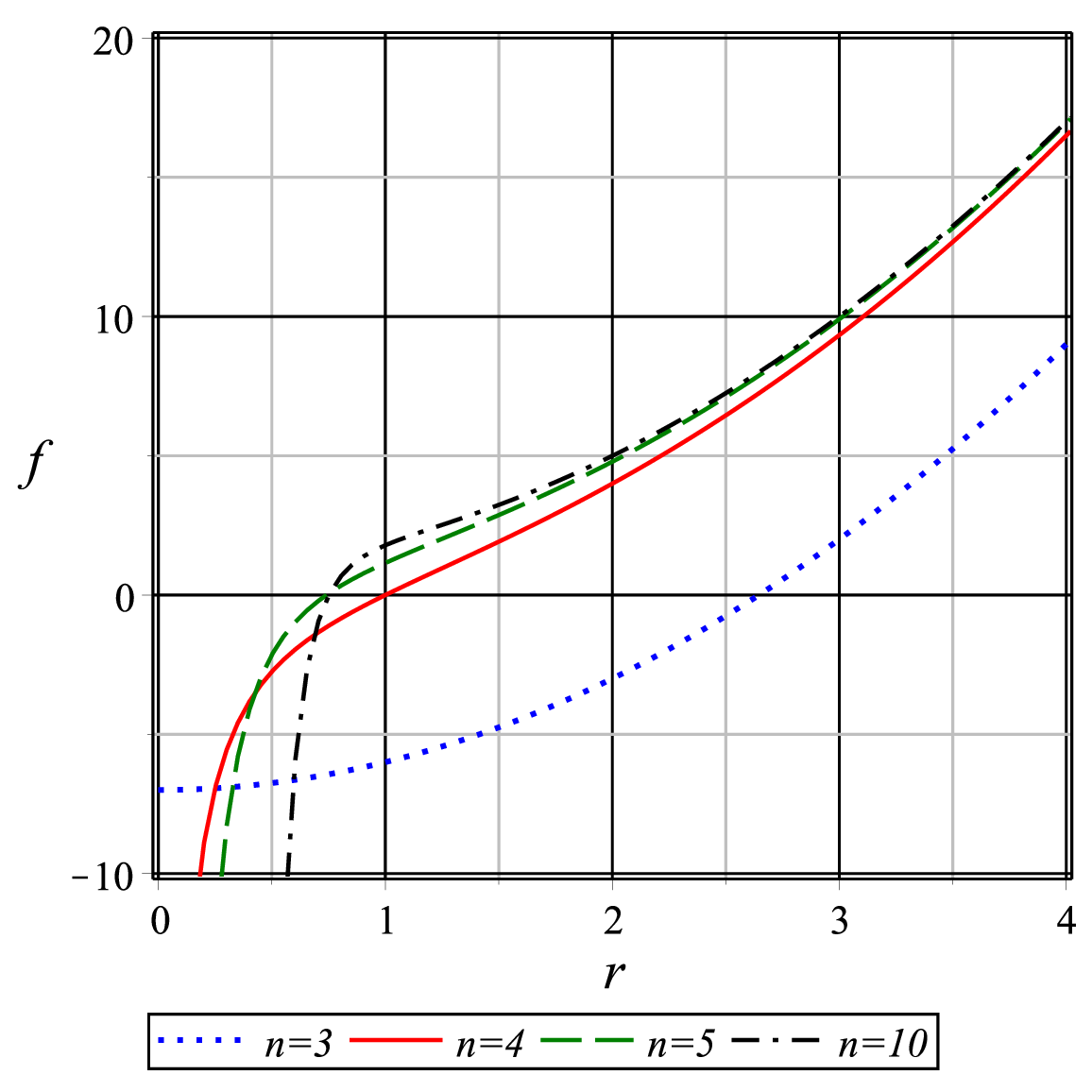}
\caption{Metric function $f(r)$ for the $n$-dimensional ST--AdS BH with $M=1$ and $l=1$.}
\label{fig-f}
\end{figure}
The Hawking temperature follows from the surface gravity:
\begin{equation}\label{T}
T=\frac{1}{4\pi}\left.\frac{df}{dr}\right|_{r=r_{h}}=\frac{(n-1)\,r_{h}^{2}+(n-3)\,l^{2}}{4\pi\, l^{2}\,r_{h}}\,,
\end{equation}
where the ADM mass has been eliminated through the horizon condition. For $n=3$, the horizon radius and temperature admit closed-form expressions:
\begin{equation}\label{rh3}
r_{h}=l\sqrt{\frac{16\pi M-\omega}{\omega}}\,,\qquad
T=\frac{1}{2\pi l}\sqrt{\frac{16\pi M-\omega}{\omega}}\,,\qquad (n=3)\,,
\end{equation}
so that $T$ grows linearly in $r_{h}$. For $n=5$, we have
\begin{equation}\label{rh5}
r_{h}=\sqrt{\frac{-l^{2}+\sqrt{l^{4}+\frac{64\pi M l^{2}}{3\omega}}}{2}}\,,\qquad (n=5)\,,
\end{equation}
with the temperature
\begin{equation}\label{T5}
T=\frac{1}{2\pi\,r_{h}}+\frac{r_{h}}{\pi\, l^{2}}\,,\qquad (n=5)\,.
\end{equation}
For $n\geq 4$, the temperature exhibits a minimum $T_{\min}$ at $r_{h}^{\min}=l\sqrt{(n-3)/(n-1)}$, below which the canonical ensemble becomes unstable. Setting $l=1$, the minimum temperatures are $T_{\min}\simeq 0.2757$ at $r_{h}^{\min}\simeq 0.5774$ ($n=4$), $T_{\min}\simeq 0.4502$ at $r_{h}^{\min}\simeq 0.7071$ ($n=5$), and $T_{\min}\simeq 1.2633$ at $r_{h}^{\min}\simeq 0.8819$ ($n=10$). These features are displayed in Fig.~\ref{fig-T}.

\begin{figure}[h!]
\centering
\includegraphics[width=120mm]{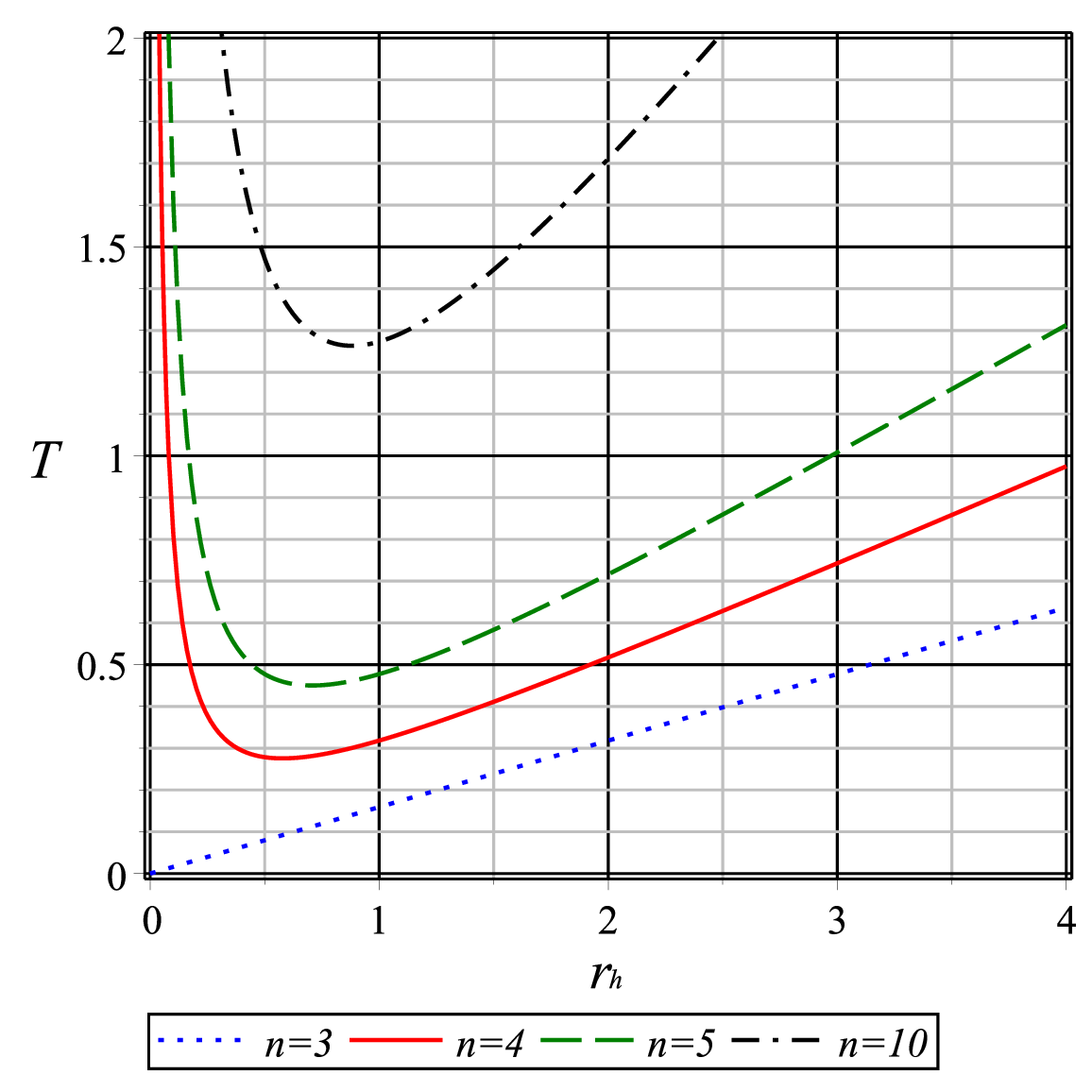}
\caption{Hawking temperature of the $n$-dimensional ST--AdS BH as a function of $r_{h}$, for $l=1$.}
\label{fig-T}
\end{figure}

\subsection{Perturbative and non-perturbative entropy corrections}\label{subsec:corrections}

Through the AdS/CFT correspondence, BH thermodynamics can be studied via the dual CFT partition function. The modular invariance of this partition function~\cite{mi12,mi14,mi16,mi18,prl,12,12a,1208.6251,1210.4617,mp9} implies that the leading quantum correction to the BH entropy takes the logarithmic form $S_{\rm per}\sim\ln S_{0}$, where $S_{0}$ is the equilibrium Bekenstein--Hawking entropy~\cite{1e,1f,1g,1h,1i,1j}. The argument proceeds as follows. One writes the entropy as a function of the inverse temperature, $S(\beta)=a\,\beta^{n}+b\,\beta^{m}$ with $a,b,n,m>0$. At the extremum $\beta_{0}=(nb/ma)^{1/(m+n)}=T^{-1}$, the equilibrium entropy is $S_{0}=S(\beta_{0})$. Small fluctuations around this saddle point produce the logarithmic correction~\cite{1e,1f,1g,1h}.

These perturbative corrections capture the physics of BHs that are small but still well above the Planck scale. Once the BH size becomes comparable to the Planck length, the perturbative expansion breaks down and non-perturbative effects dominate. It has been argued on general grounds~\cite{3}, and confirmed through supergravity functional integrals~\cite{3aa,ds12,ds14}, that the non-perturbative correction to the entropy takes the universal exponential form
\begin{equation}\label{nonper}
S_{\rm non\text{-}per}\sim e^{-S_{0}}\,.
\end{equation}
This structure has been applied to spherically symmetric BHs~\cite{3cd}, Born--Infeld BHs in a cavity~\cite{3bb}, and black branes~\cite{3ab}, consistently modifying the thermodynamic stability and phase structure at small scales.

Following Refs.~\cite{1i,1j}, we introduce a dimensionless control parameter $\eta$ to track the strength of the non-perturbative correction, and write the total corrected entropy as
\begin{equation}\label{Scorr}
S=S_{0}+\eta\,e^{-S_{0}}\,.
\end{equation}
For the $n$-dimensional ST--AdS BH, the Bekenstein--Hawking entropy is $S_{0}=\frac{\omega}{2}\,r_{h}^{n-2}$, so that
\begin{equation}\label{mod-ent}
S=\frac{\omega}{2}\,r_{h}^{n-2}+\eta\,\exp\!\left(-\frac{\omega}{2}\,r_{h}^{n-2}\right).
\end{equation}
The exponential term is negligible for large $r_{h}$ (i.e.\ when $S_{0}\gg 1$), recovering the standard area law. It becomes significant only at very small horizon radii, where the BH probes the Planck regime. The parameter $\eta$ interpolates between the uncorrected case ($\eta=0$) and the fully corrected one. We note that the corrections are formulated in terms of the original equilibrium quantities $S_{0}$ and $T$; the temperature~\eqref{T} is therefore not modified.

In Fig.~\ref{fig-entropy}, we plot the corrected entropy as a function of $r_{h}$ for $n=4$ and $l=1$, comparing $\eta=0$ (Bekenstein--Hawking) and $\eta=1$ (fully corrected). The two curves are virtually indistinguishable for $r_{h}\gtrsim 1$, where $S_{0}$ is already large enough to suppress the exponential correction. At small $r_{h}$, the $\eta=1$ curve deviates upward: the non-perturbative term $e^{-S_{0}}$ adds a positive contribution that prevents the entropy from vanishing as $r_{h}\to 0$, effectively introducing a quantum ``floor'' for the entropy.

\begin{figure}[h!]
\centering
\includegraphics[width=120mm]{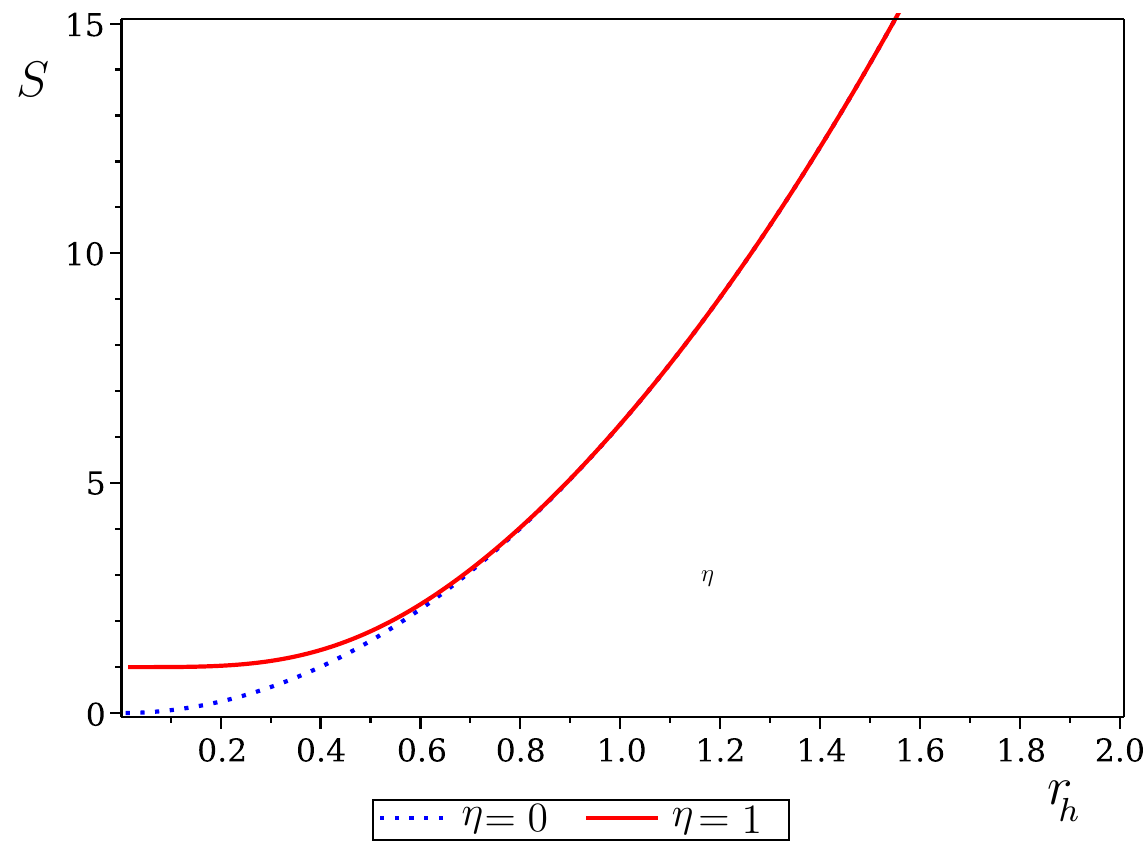}
\caption{Corrected entropy $S$ as a function of $r_{h}$ for the $n=4$ ST--AdS BH with $l=1$. The dotted blue curve is the uncorrected Bekenstein--Hawking entropy ($\eta=0$), and the solid red curve includes the non-perturbative correction ($\eta=1$). The deviation is pronounced only at small horizon radii.}
\label{fig-entropy}
\end{figure}

The quantitative impact of the correction is shown in Table~\ref{tab:Scorr}. For $r_{h}=0.1$, the relative correction $\Delta S/S_{0}=\eta\,e^{-S_{0}}/S_{0}$ exceeds $1490\%$, demonstrating that the non-perturbative term dominates in this regime. By $r_{h}=0.5$, the correction drops to about $13\%$, and for $r_{h}\geq 1$ it falls below $0.03\%$, becoming entirely negligible. This confirms that the exponential correction selectively targets the small-BH (quantum) regime without altering the thermodynamics of macroscopic BHs.

\begin{table}[h!]
\centering
\renewcommand{\arraystretch}{1.5}
\caption{Corrected entropy $S$ from Eq.~\eqref{mod-ent} for selected $r_{h}$ values, with $n=4$ and $l=1$. The column $\Delta S/S_{0}$ gives the relative correction for $\eta=1$.}\label{tab:Scorr}
\begin{tabular}{c c c c c}
\hline\hline
$r_{h}$ & $S_{0}$ & $S\;(\eta=0)$ & $S\;(\eta=1)$ & $\Delta S/S_{0}$ \\
\hline
$0.1$ & $0.0628$ & $0.0628$ & $1.0019$ & $14.9463$ \\
$0.3$ & $0.5655$ & $0.5655$ & $1.1336$ & $1.0046$ \\
$0.5$ & $1.5708$ & $1.5708$ & $1.7787$ & $0.1323$ \\
$1.0$ & $6.2832$ & $6.2832$ & $6.2851$ & $0.0003$ \\
$2.0$ & $25.133$ & $25.133$ & $25.133$ & $<10^{-6}$ \\
$5.0$ & $157.08$ & $157.08$ & $157.08$ & $<10^{-6}$ \\
\hline\hline
\end{tabular}
\end{table}

\section{Thermodynamic stability and corrected thermodynamics}\label{isec3}

In this section, we examine the effect of the non-perturbative quantum gravitational corrections on the thermodynamic stability of the ST--AdS BH, and we derive the corrected internal energy, Helmholtz free energy, and Gibbs free energy. Throughout, the corrected entropy $S$ is taken from Eq.~\eqref{mod-ent}.

\subsection{Specific heat and local stability}\label{subsec:CV}

The local thermodynamic stability of a BH is governed by the sign of the specific heat at constant volume (i.e.\ at fixed AdS radius $l$),
\begin{equation}\label{Cv}
C_{V}=T\left(\frac{\partial S}{\partial T}\right)_{V}\,.
\end{equation}
Using the corrected entropy~\eqref{mod-ent} together with the Hawking temperature~\eqref{T}, the quantum-corrected specific heat of the $n$-dimensional ST--AdS BH can be written explicitly as
\begin{equation}\label{CVexplicit}
C_{V}=\left(\frac{n-3}{4\pi r_{h}}+\frac{(n-1)\,r_{h}}{4\pi l^{2}}\right)\left(\frac{\frac{(n-2)\,\omega}{2}\,r_{h}^{n-3}}{-\frac{n-3}{4\pi r_{h}^{2}}+\frac{n-1}{4\pi l^{2}}}\right)\left(1-\eta\,e^{-S_{0}}\right).
\end{equation}
The denominator vanishes at
\begin{equation}\label{rhstar}
r_{h}^{*}=l\sqrt{\frac{n-3}{n-1}}\,,
\end{equation}
which coincides with the minimum of the Hawking temperature identified in Sec.~\ref{subsec:STAdS}. At this point $C_{V}$ diverges, signalling a second-order phase transition between two branches: a thermodynamically stable large-BH branch ($r_{h}>r_{h}^{*}$, $C_{V}>0$) and an unstable small-BH branch ($r_{h}<r_{h}^{*}$, $C_{V}<0$). Hence, for $n\geq 4$, the BH undergoes a phase transition and becomes thermodynamically unstable at sufficiently small scales.

The factor $(1-\eta\,e^{-S_{0}})$ encodes the non-perturbative quantum correction. For large BHs ($S_{0}\gg 1$), this factor is indistinguishable from unity and the standard result is recovered. For instance, at $r_{h}=2$ and $n=4$ ($l=1$), we find $C_{V}\approx 59.40$ for both $\eta=0$ and $\eta=1$, confirming that the correction is negligible for macroscopic BHs. For quantum-sized BHs, however, $e^{-S_{0}}$ becomes appreciable and noticeably reduces the magnitude of $C_{V}$: at $r_{h}=0.2$, $C_{V}$ drops from $-0.64$ ($\eta=0$) to $-0.14$ ($\eta=1$), a reduction by nearly $78\%$. In the extreme case $\eta\,e^{-S_{0}}\to 1$, the specific heat is driven toward zero, indicating a quantum suppression of the thermal response near the Planck scale.

The $n=3$ case is special: $(n-3)=0$ removes the divergence, and $C_{V}$ remains positive for all $r_{h}$. The three-dimensional AdS BH is therefore thermodynamically stable regardless of its size, though the quantum correction still reduces $C_{V}$ at small horizon radii. In Fig.~\ref{figCv}, we display $C_{V}$ as a function of $r_{h}$ for $n=3,4,5,10$. For $n=4$, the divergence occurs at $r_{h}^{*}=l/\sqrt{3}\approx 0.577$: numerically, $C_{V}(\eta=0)$ jumps from $-318.7$ at $r_{h}=0.57$ to $+923.2$ at $r_{h}=0.58$, confirming the sign change. The corrected values ($\eta=1$) follow the same pattern with reduced magnitude ($-277.3$ and $+811.7$, respectively). The $n=3$ panel confirms the absence of a phase transition, while the remaining panels show the characteristic divergence at $r_{h}^{*}$ and the suppression of $C_{V}$ on the small-BH branch.

\begin{figure}[h!]
\begin{center}$
\begin{array}{cc}
\includegraphics[width=80mm]{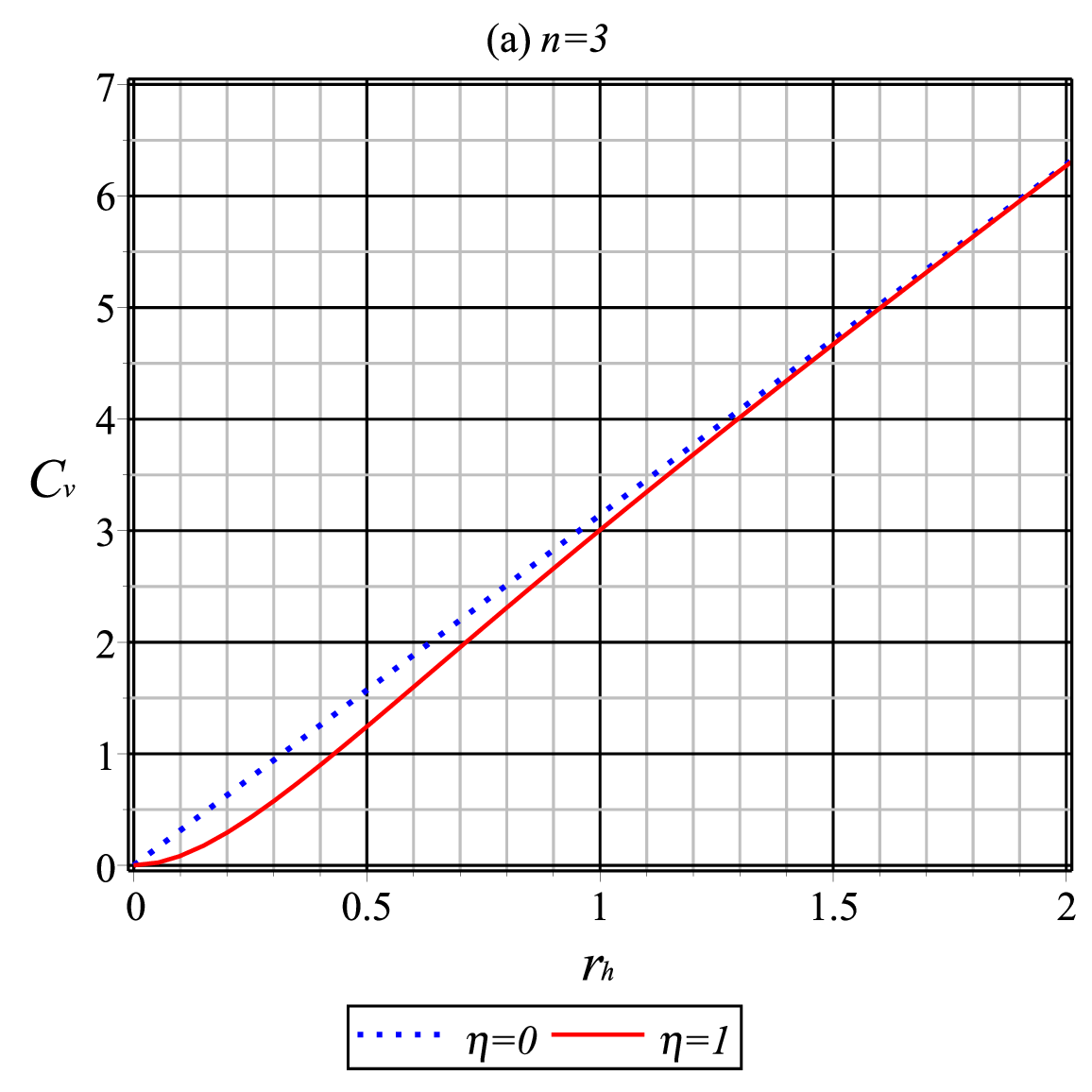} & \includegraphics[width=80mm]{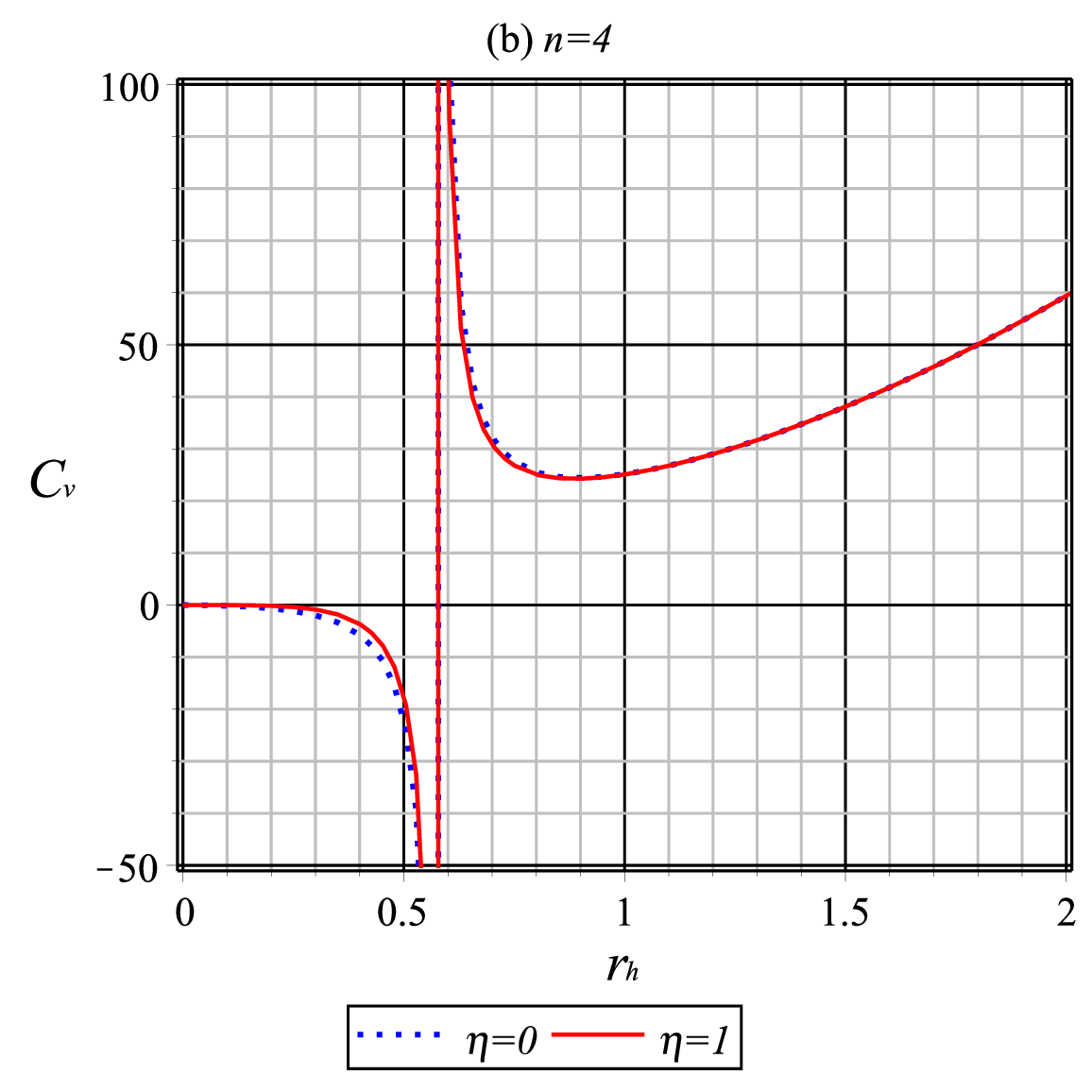}\\
\includegraphics[width=80mm]{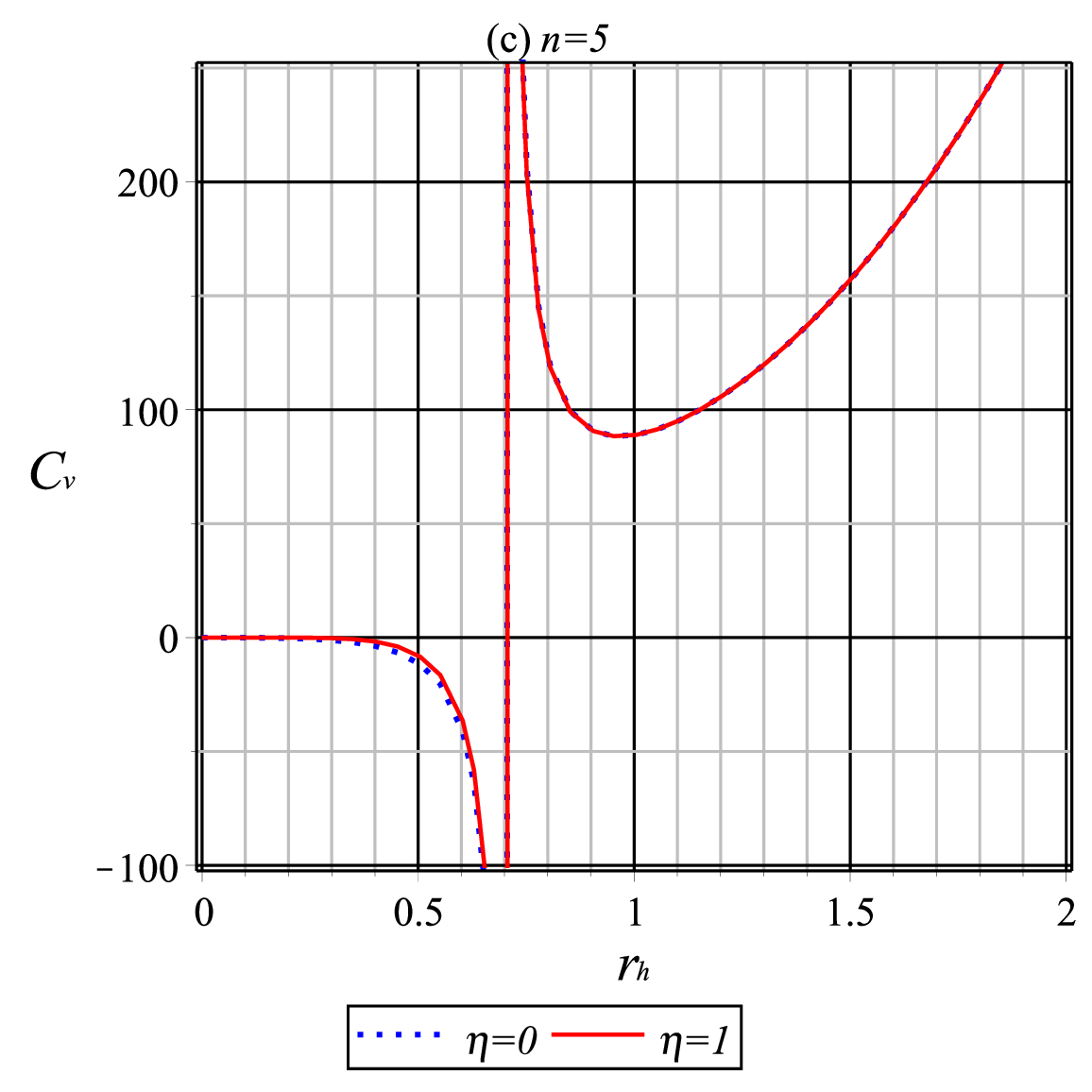} & \includegraphics[width=80mm]{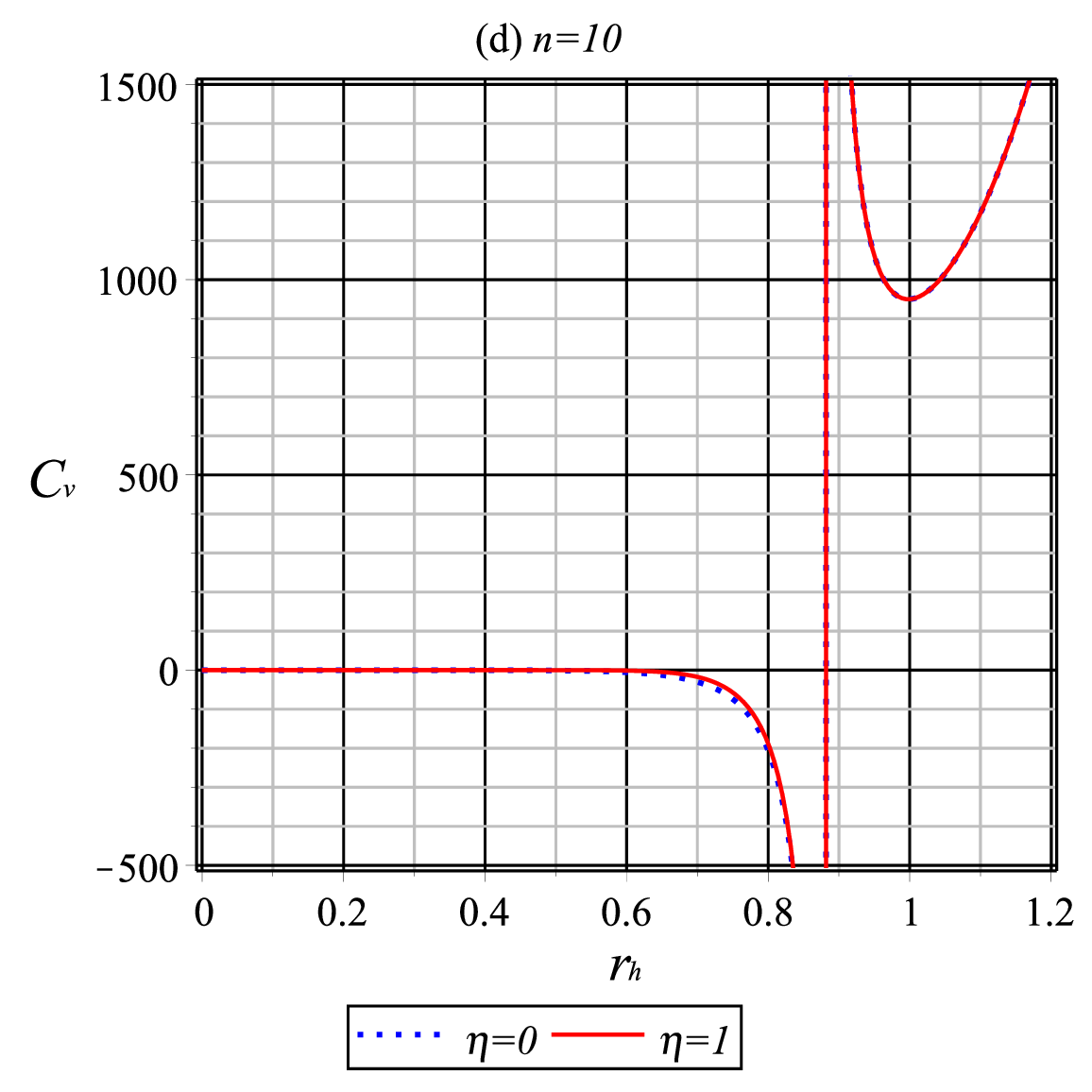}
\end{array}$
\end{center}
\caption{Specific heat $C_{V}$ of the $n$-dimensional ST--AdS BH for $l=1$. Panels correspond to $n=3$ (top left), $n=4$ (top right), $n=5$ (bottom left), and $n=10$ (bottom right). Solid curves: $\eta=0$; dashed curves: $\eta=1$. The divergence at $r_{h}^{*}=l\sqrt{(n-3)/(n-1)}$ marks the phase transition for $n\geq 4$.}
\label{figCv}
\end{figure}

\subsection{Extended phase space}\label{subsec:PV}

For an AdS BH, the cosmological constant can be treated as a thermodynamic pressure, and a volume conjugate to it can be used to construct an extended phase space~\cite{pv1,pv2,pv4,pv5}. The pressure and volume of the ST--AdS BH are given by~\cite{5,51,52,54}
\begin{equation}\label{PV}
P=\frac{(n-1)(n-2)}{16\pi l^{2}}\,,\qquad V=\frac{\omega}{n-1}\,r_{h}^{n-1}\,.
\end{equation}
Introducing the specific volume
\begin{equation}\label{specvol}
v=\frac{4}{n-2}\left(\frac{(n-1)\,V}{\omega_{n-2}}\right)^{\!\frac{1}{n-1}},
\end{equation}
the equation of state takes the form~\cite{12a}
\begin{equation}\label{EoS}
\frac{Pv}{T}=1-\frac{n-3}{(n-2)\pi}\,\frac{1}{Tv}\,.
\end{equation}
Comparing with the virial expansion
\begin{equation}\label{virial}
\frac{Pv}{T}=1+\frac{B(T)}{v}+\frac{C(T)}{v^{2}}+\frac{D(T)}{v^{3}}+\cdots\,,
\end{equation}
we identify the sole non-vanishing virial coefficient as
\begin{equation}\label{B}
B(T)=-\frac{n-3}{(n-2)\pi\, T}\,,
\end{equation}
with $C(T)=D(T)=\cdots=0$. (For a four-dimensional charged BH, $D(T)\neq 0$.) The Boyle temperature, at which $B(T)$ vanishes, is formally infinite for the uncharged ST--AdS case. The equation of state~\eqref{EoS} is a special case of the vdW equation
\begin{equation}\label{vdW}
Pv=\frac{T}{v-b}+\frac{a}{v^{2}}\,,
\end{equation}
with $b=0$ and $a=(n-3)/[(n-2)\pi]$. Numerically, $a=0$ for $n=3$ (no attractive interaction in the equation of state), $a\approx 0.159$ for $n=4$, $a\approx 0.212$ for $n=5$, and $a\approx 0.279$ for $n=10$; the interaction parameter grows with $n$ but remains bounded.

For a quantum-sized BH, the corrected entropy becomes
\begin{equation}\label{Sapprox}
S\approx\frac{\omega}{2}(1-\eta)\left(\frac{n-2}{4}\,v\right)^{n-2}.
\end{equation}
Applying the critical-point conditions
\begin{equation}\label{cond}
\frac{\partial P}{\partial v}=0\,,\qquad \frac{\partial^{2}P}{\partial v^{2}}=0\,,
\end{equation}
to the equation of state~\eqref{EoS}, one finds that no critical point exists at finite $v$; the only solution is $v=0$. This absence of a vdW-type critical point is consistent with the uncharged nature of the ST--AdS BH: the relevant phase transition is of Hawking--Page type (thermal AdS $\leftrightarrow$ large BH) rather than the small-BH/large-BH liquid--gas transition that requires a non-zero charge~\cite{pv1,pv2}.

\subsection{Internal energy}\label{subsec:E}

The quantum-corrected internal energy is obtained from $E=\int T\,dS$. Carrying out the integration with the corrected entropy~\eqref{mod-ent}, we find
\begin{eqnarray}\label{Eexact}
E&=&\frac{(n-2)\,\omega}{8\pi}\left(r_{h}^{n-3}+\frac{r_{h}^{n-1}}{l^{2}}\right)+\eta\,\frac{n-3}{8\pi}\,2^{\frac{n-3}{n-2}}\,\omega^{\frac{1}{n-2}}\,\Gamma\!\left(\frac{n-3}{n-2},\frac{\omega\, r_{h}^{n-2}}{2}\right)\nonumber\\
&&+\;\eta\,\frac{n-1}{8\pi\, l^{2}}\,2^{\frac{n-1}{n-2}}\,\omega^{-\frac{1}{n-2}}\,\Gamma\!\left(\frac{n-1}{n-2},\frac{\omega\, r_{h}^{n-2}}{2}\right),
\end{eqnarray}
where $\Gamma(z,x)$ is the upper incomplete gamma function,
\begin{equation}\label{incgamma}
\Gamma(z,x)=x^{z}\,e^{-x}\sum_{k=0}^{\infty}\frac{L_{k}^{(z)}(x)}{k+1}\,,
\end{equation}
and $L_{k}^{(z)}(x)$ are the generalized Laguerre polynomials, defined through the recurrence
\begin{eqnarray}\label{Laguerre}
L_{0}^{(z)}(x)&=&1\,,\nonumber\\
L_{1}^{(z)}(x)&=&1+z-x\,,\nonumber\\
L_{k+1}^{(z)}(x)&=&\frac{(2k+1+z-x)\,L_{k}^{(z)}(x)-(k+z)\,L_{k-1}^{(z)}(x)}{k+1}\,,\qquad k\geq 1\,.
\end{eqnarray}
The first term in Eq.~\eqref{Eexact} is the standard ADM mass $M$, and the remaining terms containing the incomplete gamma functions encode the non-perturbative quantum correction. For large $r_{h}$, the asymptotic behavior $\Gamma(z,x)\sim x^{z-1}\,e^{-x}$ as $x\to\infty$ ensures that the correction is exponentially suppressed, and $E\to M$.

In Fig.~\ref{figE}, we display the internal energy as a function of $r_{h}$ for $n=3,4,5,10$. The energy increases monotonically with $r_{h}$, as expected. For $n=4$ and $l=1$, the uncorrected internal energy coincides with the ADM mass, $E(\eta=0)=M$. For instance, $E=0.101$ at $r_{h}=0.1$, rising to $E=130.0$ at $r_{h}=5$. The quantum correction is negligible for large BHs ($E(\eta=1)\approx E(\eta=0)$ for $r_{h}\geq 1$), but it raises the internal energy at small scales: at $r_{h}=0.1$, the corrected value $E(\eta=1)=0.440$ exceeds the uncorrected one by more than a factor of four. This enhancement arises from the incomplete gamma function terms in Eq.~\eqref{Eexact}, which contribute a positive, exponentially sensitive correction when $S_{0}\lesssim 1$.

\begin{figure}[h!]
\begin{center}$
\begin{array}{cc}
\includegraphics[width=80mm]{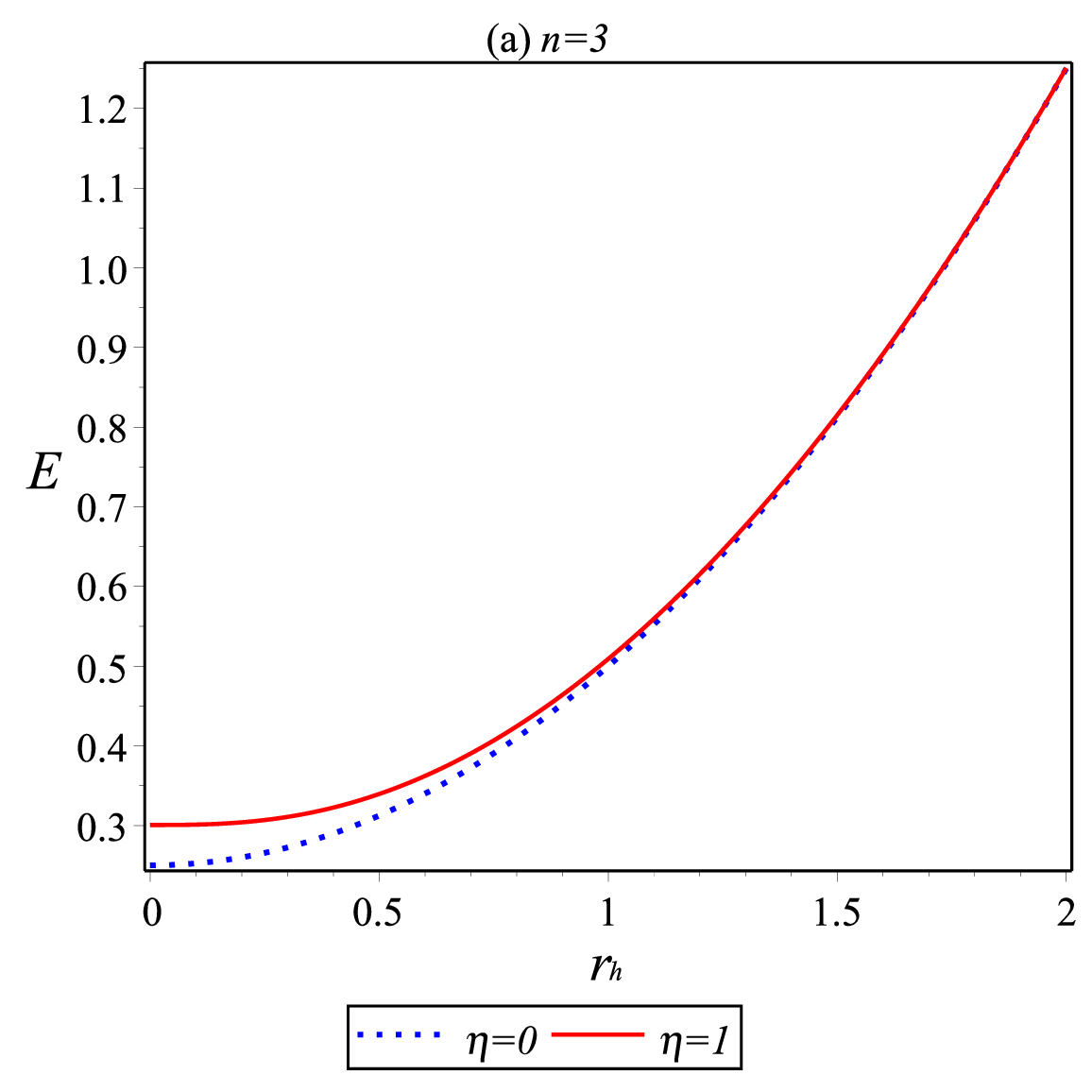} & \includegraphics[width=80mm]{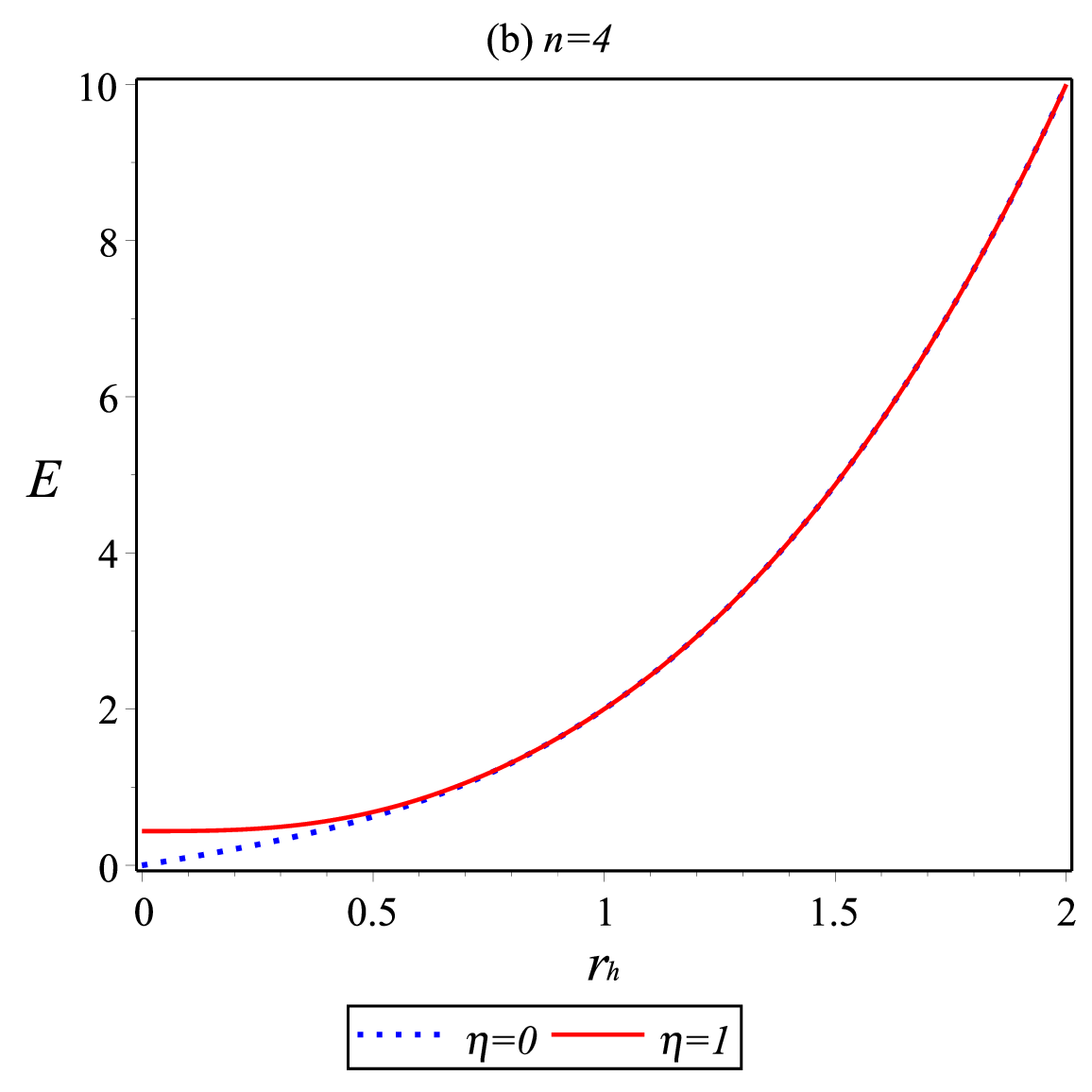}\\
\includegraphics[width=80mm]{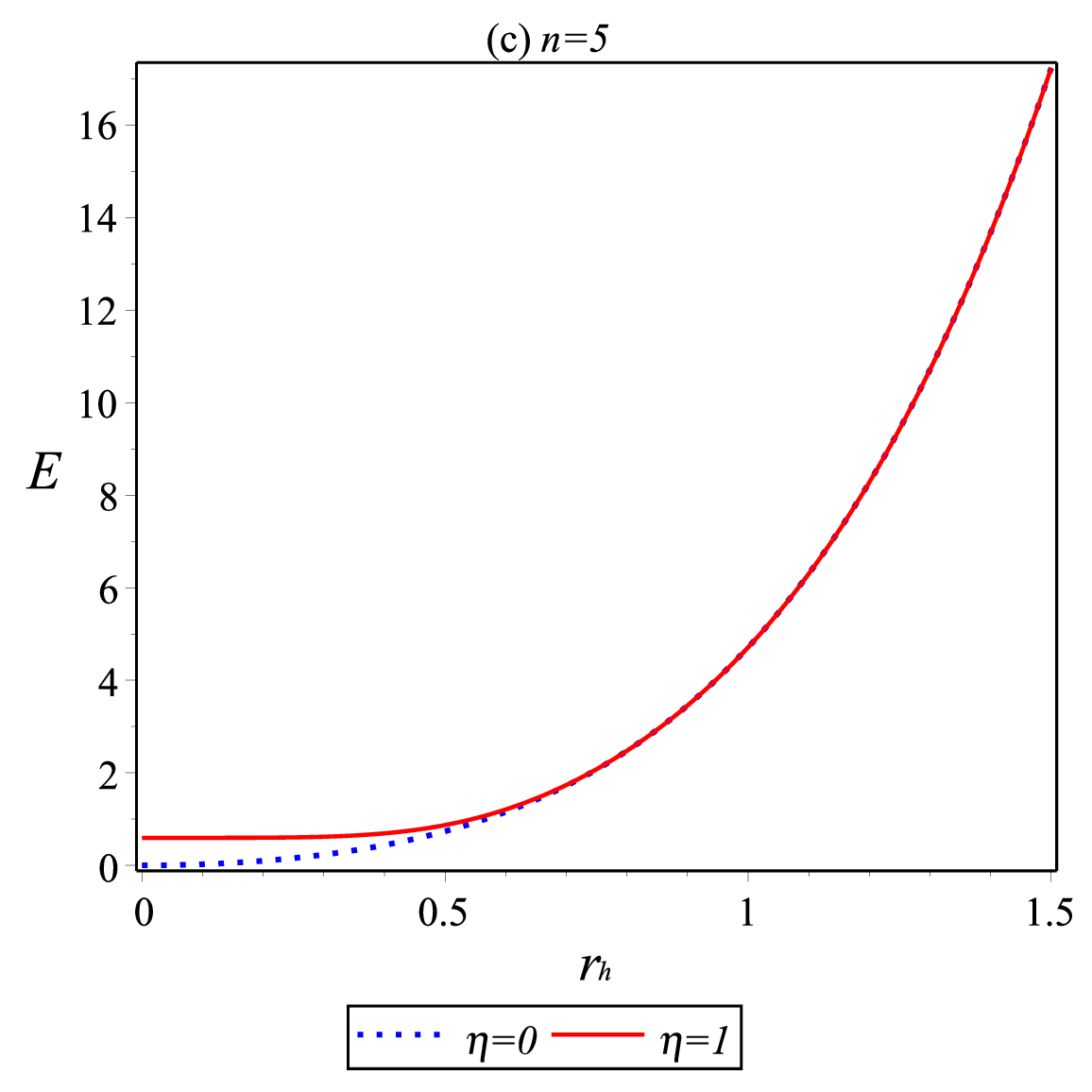} & \includegraphics[width=80mm]{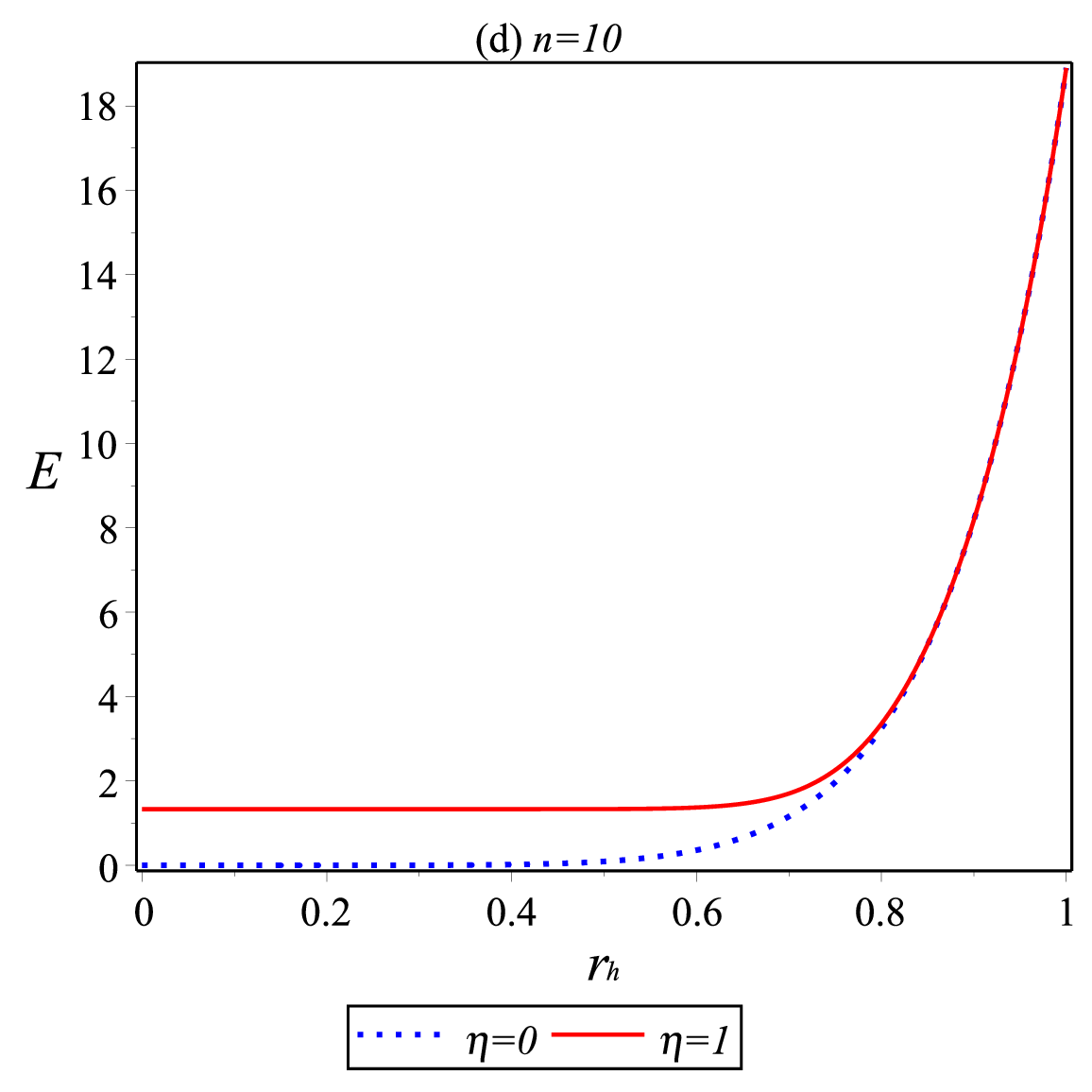}
\end{array}$
\end{center}
\caption{Internal energy $E$ of the $n$-dimensional ST--AdS BH for $l=1$. Panels correspond to $n=3$ (top left), $n=4$ (top right), $n=5$ (bottom left), and $n=10$ (bottom right). Solid curves: $\eta=0$; dashed curves: $\eta=1$.}
\label{figE}
\end{figure}

\subsection{Helmholtz and Gibbs free energies}\label{subsec:FG}

The quantum-corrected Helmholtz free energy $F=E-TS$ is obtained by combining the corrected internal energy~\eqref{Eexact} with the corrected entropy~\eqref{mod-ent} and the Hawking temperature~\eqref{T}:
\begin{eqnarray}\label{Ffull}
F&=&\frac{\omega\left[(n-2)\,r_{h}\bigl(l^{2}\,r_{h}^{n-3}+r_{h}^{n-1}\bigr)-r_{h}^{n-2}\bigl(l^{2}(n-3)+(n-1)\,r_{h}^{2}\bigr)\right]}{8\pi\, r_{h}\,l^{2}}\nonumber\\
&&+\;\frac{\eta}{8\pi\, l^{2}}\left(2^{\frac{n-1}{n-2}}\,\omega^{-\frac{1}{n-2}}(n-1)\,\Gamma\!\left(\frac{n-1}{n-2},\frac{\omega\, r_{h}^{n-2}}{2}\right)+l^{2}\,2^{\frac{n-3}{n-2}}\,\omega^{\frac{1}{n-2}}(n-3)\,\Gamma\!\left(\frac{n-3}{n-2},\frac{\omega\, r_{h}^{n-2}}{2}\right)\right)\nonumber\\
&&-\;\frac{\eta}{4\pi\, r_{h}\,l^{2}}\left(l^{2}(n-3)+(n-1)\,r_{h}^{2}\right)\exp\!\left(-\frac{\omega\, r_{h}^{n-2}}{2}\right).
\end{eqnarray}
The Helmholtz free energy determines the thermodynamic preference in the canonical ensemble. A BH configuration with $F<0$ is thermodynamically favored over thermal AdS, while $F>0$ indicates the opposite. The Hawking--Page transition occurs at the temperature where $F$ changes sign~\cite{5a,mp7}. The quantum correction, contained in the incomplete gamma functions and the last exponential term, shifts this zero-crossing to a slightly different horizon radius. The shift is negligible for large BHs but becomes appreciable for quantum-sized ones.

The Gibbs free energy in the extended phase space, $G=F+PV$, takes the form
\begin{eqnarray}\label{Gfull}
G&=&\frac{(n-2)\,\omega}{8\pi}\left(r_{h}^{n-3}+\frac{3\,r_{h}^{n-1}}{l^{2}}\right)-\frac{\omega}{8\pi\, l^{2}}\left((n-1)\,r_{h}^{2}+(n-3)\,l^{2}\right)r_{h}^{n-3}\nonumber\\
&&+\;\frac{\eta}{8\pi\, l^{2}}\left(2^{\frac{n-1}{n-2}}\,\omega^{-\frac{1}{n-2}}(n-1)\,\Gamma\!\left(\frac{n-1}{n-2},\frac{\omega\, r_{h}^{n-2}}{2}\right)+l^{2}\,2^{\frac{n-3}{n-2}}\,\omega^{\frac{1}{n-2}}(n-3)\,\Gamma\!\left(\frac{n-3}{n-2},\frac{\omega\, r_{h}^{n-2}}{2}\right)\right)\nonumber\\
&&-\;\frac{\eta}{4\pi\, r_{h}\,l^{2}}\left(l^{2}(n-3)+(n-1)\,r_{h}^{2}\right)\exp\!\left(-\frac{\omega\, r_{h}^{n-2}}{2}\right).
\end{eqnarray}
The Gibbs free energy determines the global stability in the isobaric ensemble. The Hawking--Page transition corresponds to $G=0$, separating configurations where thermal AdS is preferred ($G>0$) from those where the BH phase dominates ($G<0$). The numerical evaluation reveals a striking dimension-dependent behavior. For $n=3$ and $\eta=0$, the Gibbs free energy changes sign at $r_{h}\approx 1.414$ (corresponding to $T_{\rm HP}\approx 0.225$ for $l=1$), and the quantum correction shifts this to $r_{h}\approx 1.416$ ($T_{\rm HP}\approx 0.225$), a relative change of only $0.12\%$.

For $n\geq 4$, however, the situation is qualitatively different. In the uncorrected case ($\eta=0$), the Gibbs free energy remains positive for all $r_{h}$: for example, $G=0.05,\,0.25,\,0.50,\,1.00,\,2.50$ at $r_{h}=0.1,\,0.5,\,1.0,\,2.0,\,5.0$ when $n=4$ and $l=1$. This means the BH phase never becomes globally preferred over thermal AdS, and no Hawking--Page transition occurs within the uncorrected thermodynamics. The non-perturbative correction changes this picture at small scales. When $\eta=1$, the exponential and incomplete gamma terms in Eq.~\eqref{Gfull} drive $G$ negative at small $r_{h}$: for $n=4$, $G(\eta=1)=-0.381$ at $r_{h}=0.1$, crossing zero near $r_{h}\approx 0.200$ ($T\approx 0.446$). This quantum-induced zero-crossing also appears for $n=5$ ($r_{h}\approx 0.312$, $T\approx 0.610$) and $n=10$ ($r_{h}\approx 0.585$, $T\approx 1.371$). The effect is a genuine consequence of the non-perturbative correction: the exponential term $\eta\,e^{-S_{0}}$ in the entropy produces a negative contribution to $G$ that overwhelms the classical terms when $S_{0}\lesssim 1$, creating a narrow window at small $r_{h}$ where the quantum-corrected BH is thermodynamically favored.

In Fig.~\ref{figG}, we display the Gibbs free energy as a function of $r_{h}$. At large scales, the quantum corrections to $G$ are negligible and $G(\eta=1)\to G(\eta=0)$ as $r_{h}$ grows. At small $r_{h}$, the $\eta=1$ curves visibly depart from the uncorrected ones, and for $n\geq 4$ the sign reversal at quantum scales is evident. The magnitude of the modification grows with $n$.

\begin{figure}[h!]
\begin{center}$
\begin{array}{cc}
\includegraphics[width=80mm]{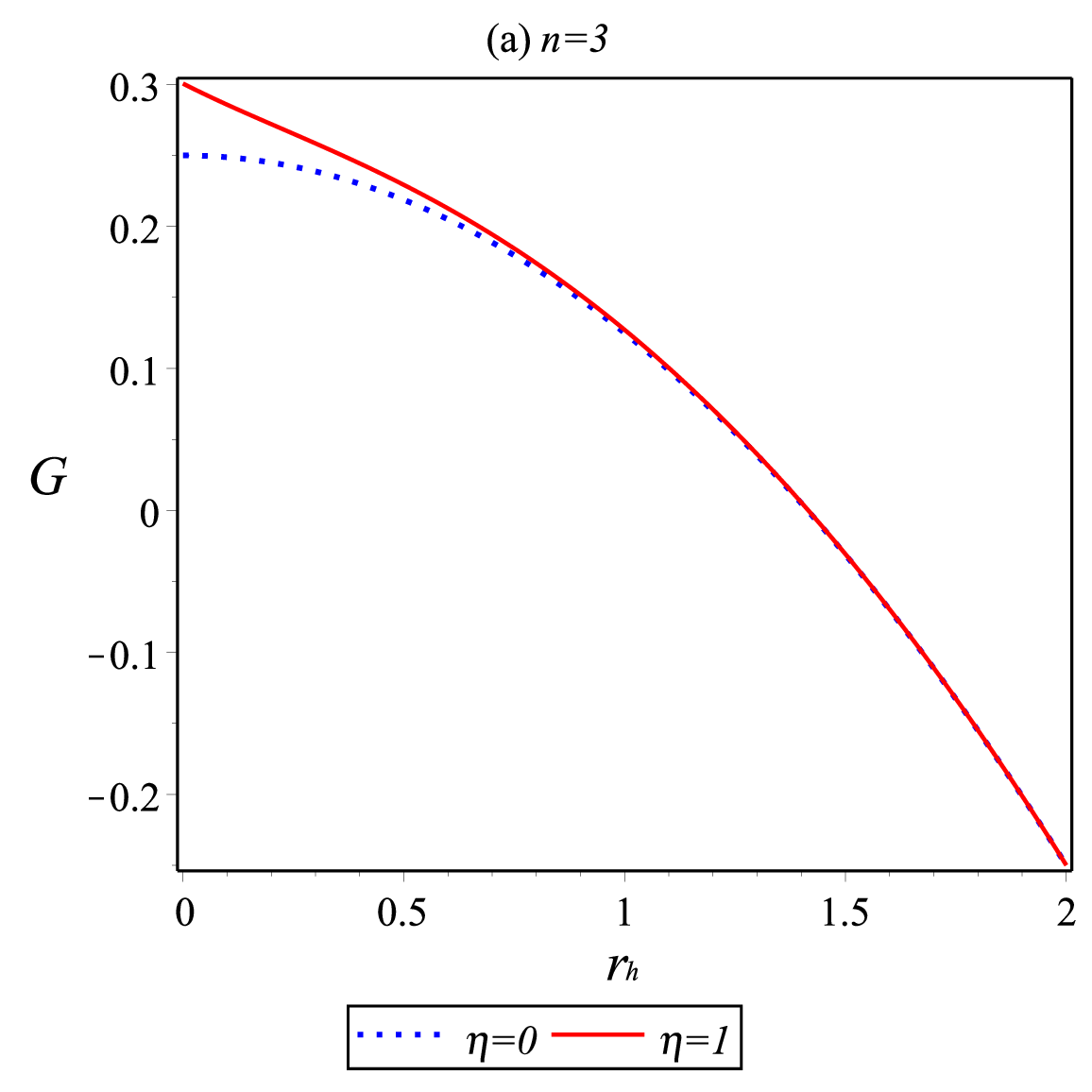} & \includegraphics[width=80mm]{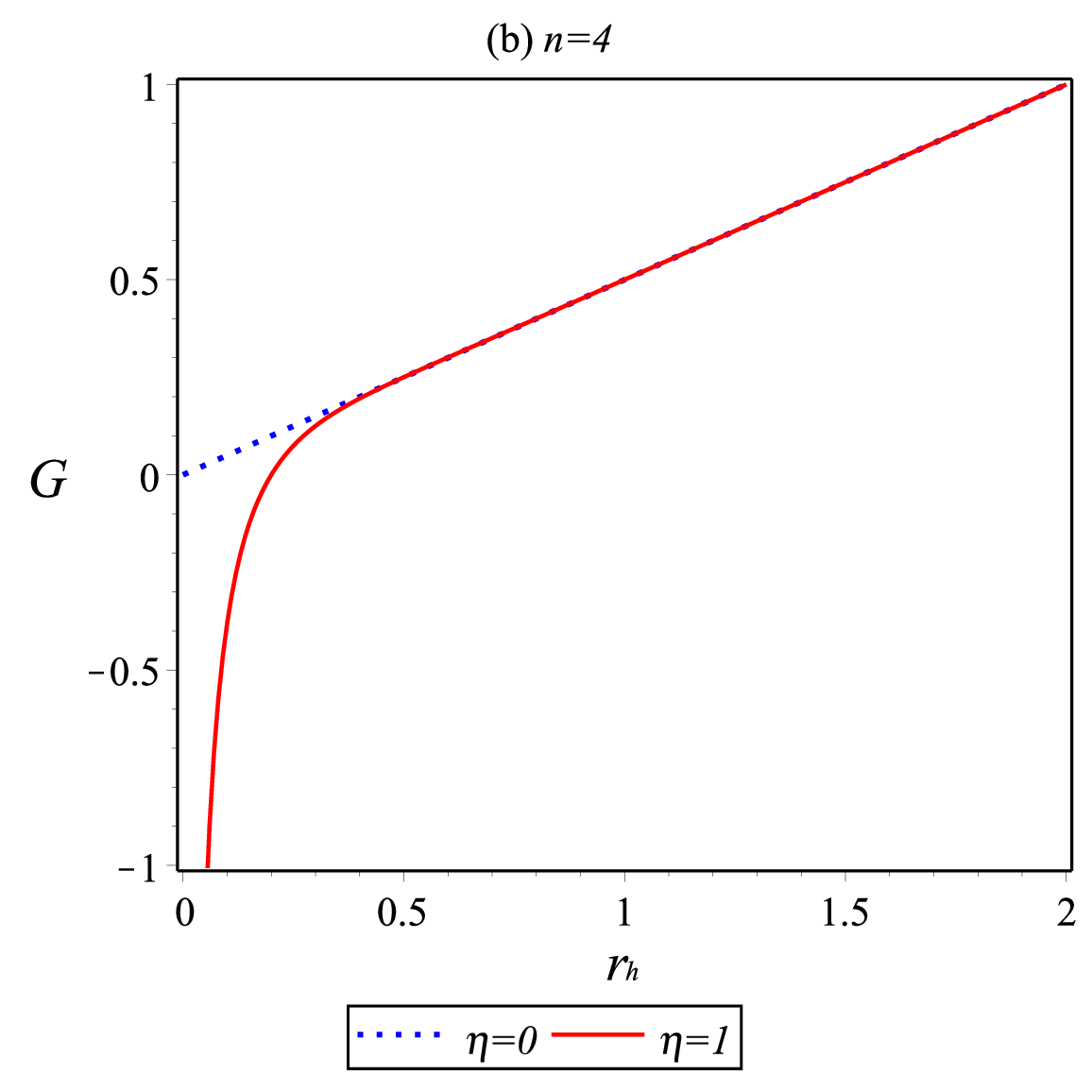}\\
\includegraphics[width=80mm]{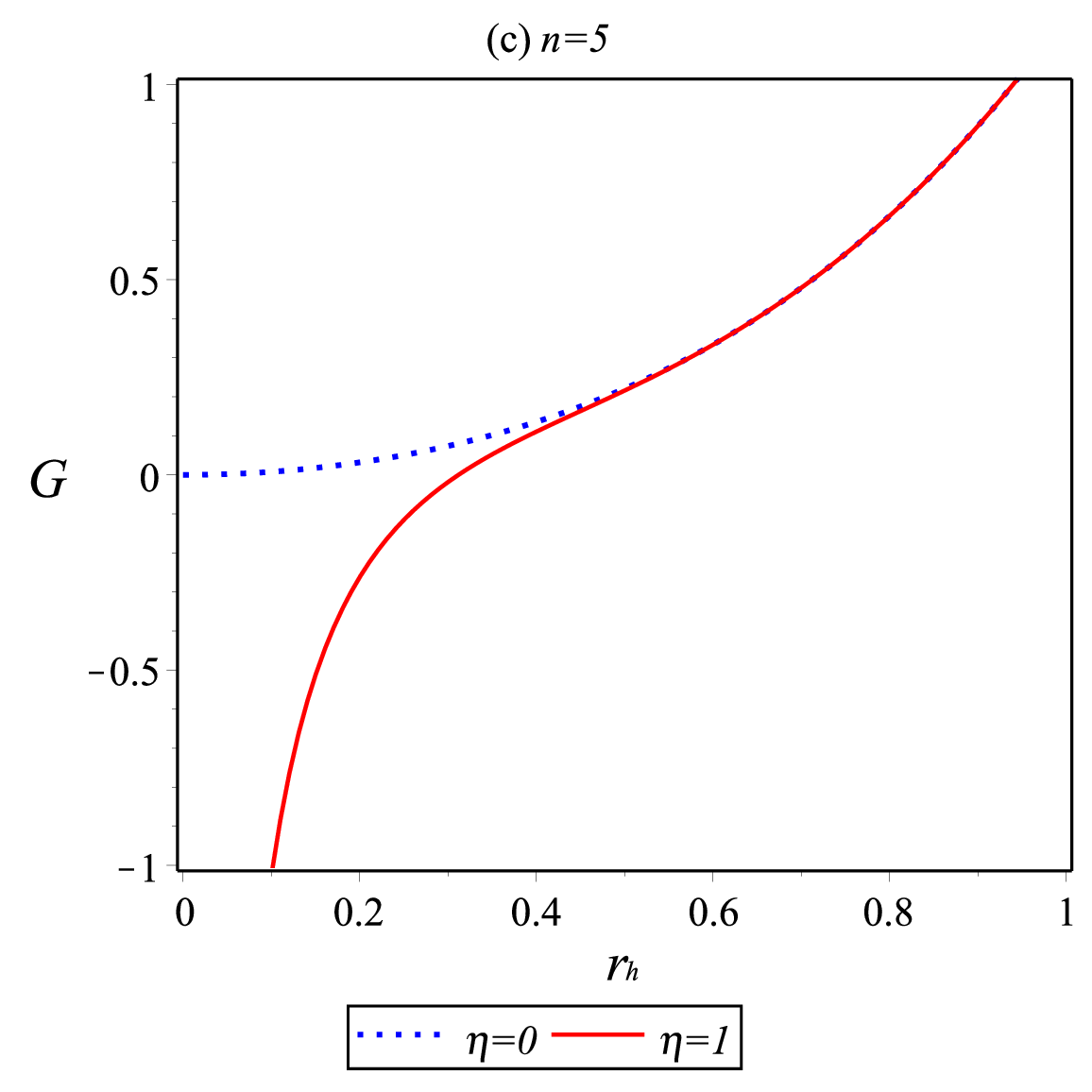} & \includegraphics[width=80mm]{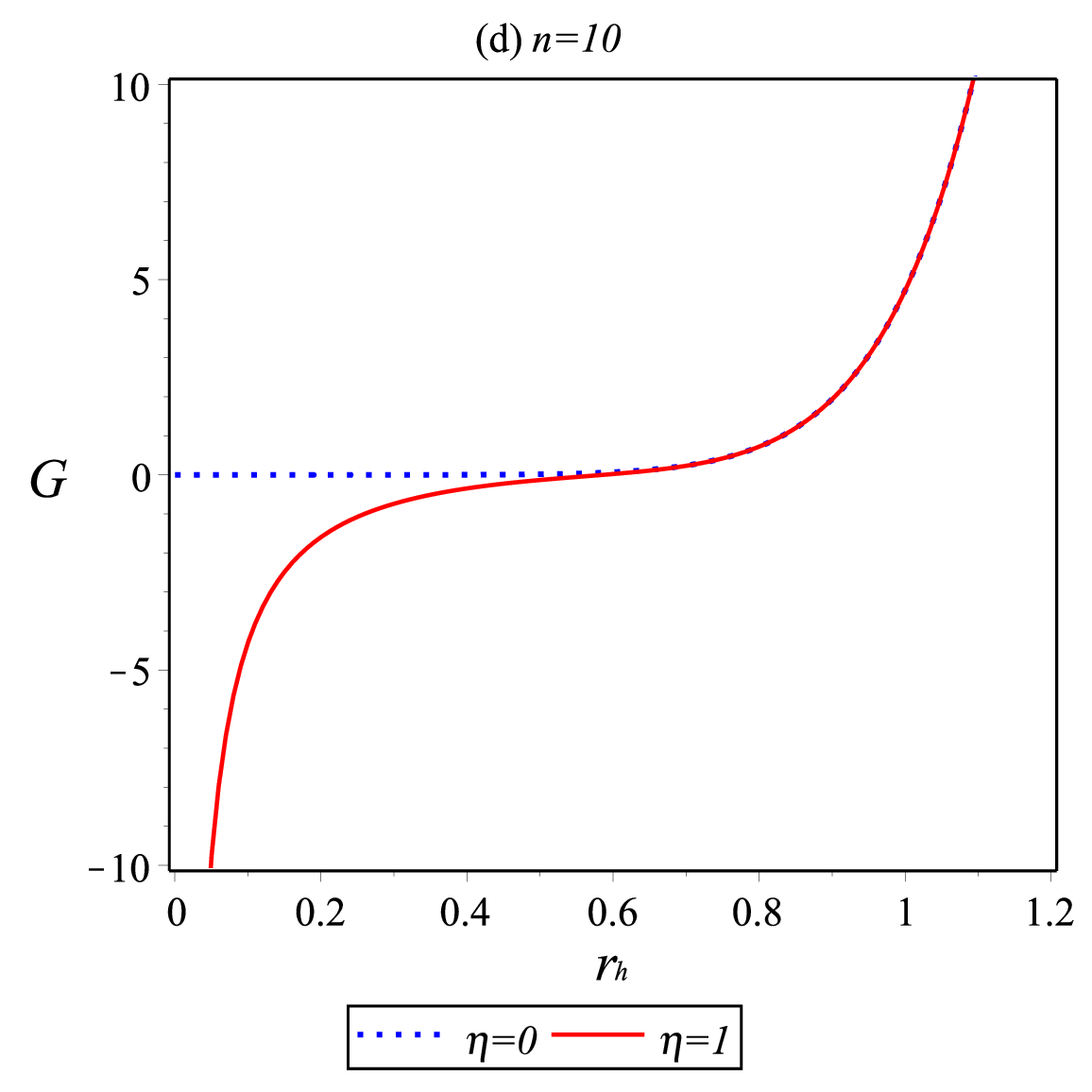}
\end{array}$
\end{center}
\caption{Gibbs free energy $G$ of the $n$-dimensional ST--AdS BH for $l=1$. Panels correspond to $n=3$ (top left), $n=4$ (top right), $n=5$ (bottom left), and $n=10$ (bottom right). Solid curves: $\eta=0$; dashed curves: $\eta=1$.}
\label{figG}
\end{figure}

In Table~\ref{tab:HP}, we collect the Hawking--Page transition data for the uncorrected and corrected cases. For $n=3$, both $\eta=0$ and $\eta=1$ yield a transition; the correction shifts $T_{\rm HP}$ upward by only $\sim 0.1\%$. For $n\geq 4$, no transition exists in the uncorrected thermodynamics, but the non-perturbative correction opens a new regime where $G<0$ at small $r_{h}$, producing a quantum-induced transition. The corresponding $T_{\rm HP}$ increases with $n$, reflecting the steeper growth of $S_{0}$ in higher dimensions.

\begin{table}[h!]
\centering
\renewcommand{\arraystretch}{1.5}
\caption{Hawking--Page transition data for the ST--AdS BH with $l=1$. The transition occurs where $G(r_{h})=0$. For $n\geq 4$ and $\eta=0$, $G>0$ for all $r_{h}$ and no transition exists.}\label{tab:HP}
\begin{tabular}{c c c c c}
\hline\hline
$n$ & $r_{h}^{\rm HP}\;(\eta=0)$ & $T_{\rm HP}\;(\eta=0)$ & $r_{h}^{\rm HP}\;(\eta=1)$ & $T_{\rm HP}\;(\eta=1)$ \\
\hline
$3$  & $1.4142$ & $0.2251$ & $1.4159$ & $0.2253$ \\
$4$  & --- & --- & $0.2000$ & $0.4457$ \\
$5$  & --- & --- & $0.3118$ & $0.6096$ \\
$10$ & --- & --- & $0.5855$ & $1.3708$ \\
\hline\hline
\end{tabular}
\end{table}

\section{Work distribution}\label{isec4}

For BHs at the Planck scale, the equilibrium description developed in the preceding sections is insufficient: the evaporation process is inherently out of equilibrium, and one must employ the formalism of non-equilibrium quantum thermodynamics~\cite{mp6,mp7}. In this section, we derive the work distribution for an evaporating ST--AdS BH, using the corrected free energies obtained in Sec.~\ref{isec3}.

Let us consider an ST--AdS BH that evaporates from a state $\Omega_{1}$ (with horizon radius $r_{h1}$) to a smaller state $\Omega_{2}$ (with horizon radius $r_{h2}<r_{h1}$). During this process, the partition function changes from $Z_{1}[\Omega_{1}]$ to $Z_{2}[\Omega_{2}]$, and the thermodynamic quantities shift accordingly. Since the quantum gravitational corrections become important only at small horizon radius, we can express the difference in the corrected entropy of a quantum-sized AdS BH as
\begin{equation}\label{DeltaS}
\Delta S=\frac{\omega}{2}(1-\eta)\left(r_{h2}^{n-2}-r_{h1}^{n-2}\right).
\end{equation}
This difference between the quantum-corrected entropies can be used to obtain the difference between the internal energy of the BH as it evaporates from $\Omega_{1}$ to $\Omega_{2}$:
\begin{eqnarray}\label{DeltaE}
\Delta E&=&\frac{(n-2)\,\omega}{8\pi}\left(r_{h2}^{n-3}-r_{h1}^{n-3}+\frac{r_{h2}^{n-1}-r_{h1}^{n-1}}{l^{2}}\right)\nonumber\\
&+&\eta\,\frac{n-3}{8\pi}\,2^{\frac{n-3}{n-2}}\,\omega^{\frac{1}{n-2}}\left[\Gamma\!\left(\frac{n-3}{n-2},\frac{1}{2}r_{h2}^{n-2}\omega\right)-\Gamma\!\left(\frac{n-3}{n-2},\frac{1}{2}r_{h1}^{n-2}\omega\right)\right]\nonumber\\
&+&\eta\,\frac{n-1}{8\pi\, l^{2}}\,2^{\frac{n-1}{n-2}}\,\omega^{-\frac{1}{n-2}}\left[\Gamma\!\left(\frac{n-1}{n-2},\frac{1}{2}r_{h2}^{n-2}\omega\right)-\Gamma\!\left(\frac{n-1}{n-2},\frac{1}{2}r_{h1}^{n-2}\omega\right)\right].
\end{eqnarray}
These corrections to the internal energy produce quantum gravitational modifications to the Hawking radiation emitted by a quantum-sized AdS BH. We denote the total heat obtained from this corrected Hawking radiation by $Q$. At such scales, the effects of work distribution cannot be neglected as the BH shrinks. Denoting the average work distribution by $\langle W\rangle$, we write
\begin{equation}\label{firstlaw}
\Delta E=Q-\langle W\rangle\,.
\end{equation}

Now let us consider the BH with partition function $Z_{1}[\Omega_{1}]$ that evaporates to a configuration with partition function $Z_{2}[\Omega_{2}]$. The ratio $Z_{2}/Z_{1}$ can be related to the quantum work distribution where states for a system starting in thermal equilibrium at temperature $T = 1/\beta$ are subjected to a process (possibly far from equilibrium) that takes it from a state $\Omega_1$ to a state $\Omega_2$, the Jarzynski equality \cite{eq12,eq14} establishes that
\begin{equation}\label{Jarzynski}
\langle \exp(-\beta W)\rangle=e^{-\beta\,\Delta F}=\frac{Z_{2}}{Z_{1}}\,.
\end{equation}
where $W$ is the work performed in each individual realization of the process and $\Delta F$ is the equilibrium free-energy difference between the final and initial states. In this sense, although each individual realization may yield a work value very different from $\Delta F$ (some with $W > \Delta F$, others with $W < \Delta F$), the exponentially weighted average of the work is exactly equal to the equilibrium free-energy difference. This is a fluctuation theorem: it connects non-equilibrium quantities (stochastic work) to an equilibrium quantity (free energy) and holds for arbitrary processes, provided the initial state is one of equilibrium. In our case, it links the quantum work distribution of black hole evaporation to free energies modified by the non-perturbative correction \cite{eq14}.

The relative weights of the partition functions can also be expressed in terms of the difference between the equilibrium free energies as $\exp(\beta\,\Delta F)=Z_{2}/Z_{1}$. Combining this with the Jarzynski equality~\cite{eq12,eq14}, the work distribution can be related to the free energy difference:
\begin{equation}\label{JarzynskiF}
\langle \exp(-\beta W)\rangle=\exp(\beta\,\Delta F)\,.
\end{equation}

We can use this quantum-corrected free energy to calculate the average work distribution between the two BH states. The Jensen inequality relates the average of the exponential of work distribution to the exponential of the average of work distribution: $\exp\langle -\beta W\rangle\leq\langle\exp(-\beta W)\rangle$, which is convex in $W$. Applying Jensen's inequality, $f(\langle x \rangle) \le \langle f(x) \rangle$, we obtain
\begin{equation}
e^{-\beta\langle W\rangle} \le \langle e^{-\beta W}\rangle = e^{-\beta\,\Delta F} \quad\Longrightarrow\quad \langle W\rangle \ge \Delta F.
\end{equation}
This is the statistical form of the second law of thermodynamics: the average work performed is always greater than or equal to the change in free energy, with equality holding only for reversible (quasi-static) processes. In the context of the black hole, it guarantees that the average quantum work $\langle W \rangle$ is bounded from below by $\Delta F$; it is precisely this relationship that, combined with the sign inversion of $\Delta F$ at small horizon radii (for $n \ge 4$ and $\eta = 1$), produces the inversion of $\langle W \rangle$ from negative to positive. Using this inequality, the quantum work distribution during the evaporation of the AdS BH can be expressed as
\begin{eqnarray}\label{Wavg}
\langle W\rangle&\approx&-\frac{\omega\left((n-2)\,r_{h2}(l^{2}\,r_{h2}^{n-3}+r_{h2}^{n-1})-r_{h2}^{n-2}(l^{2}(n-3)+(n-1)\,r_{h2}^{2})\right)}{8\pi\, r_{h2}\,l^{2}}\nonumber\\
&+&\frac{\omega\left((n-2)\,r_{h1}(l^{2}\,r_{h1}^{n-3}+r_{h1}^{n-1})-r_{h1}^{n-2}(l^{2}(n-3)+(n-1)\,r_{h1}^{2})\right)}{8\pi\, r_{h1}\,l^{2}}\nonumber\\
&-&\frac{\eta}{8\pi\, l^{2}}\left(2^{\frac{n-1}{n-2}}\,\omega^{-\frac{1}{n-2}}(n-1)\,\Gamma\!\left(\frac{n-1}{n-2},\frac{1}{2}r_{h2}^{n-2}\omega\right)+l^{2}\,2^{\frac{n-3}{n-2}}\,\omega^{\frac{1}{n-2}}(n-3)\,\Gamma\!\left(\frac{n-3}{n-2},\frac{1}{2}r_{h2}^{n-2}\omega\right)\right)\nonumber\\
&+&\frac{\eta}{8\pi\, l^{2}}\left(2^{\frac{n-1}{n-2}}\,\omega^{-\frac{1}{n-2}}(n-1)\,\Gamma\!\left(\frac{n-1}{n-2},\frac{1}{2}r_{h1}^{n-2}\omega\right)+l^{2}\,2^{\frac{n-3}{n-2}}\,\omega^{\frac{1}{n-2}}(n-3)\,\Gamma\!\left(\frac{n-3}{n-2},\frac{1}{2}r_{h1}^{n-2}\omega\right)\right)\nonumber\\
&-&\frac{2\eta}{8\pi\, r_{h2}\,l^{2}}\left(l^{2}(n-3)+(n-1)\,r_{h2}^{2}\right)\left(1-\frac{\omega\, r_{h2}^{n-2}}{2}\right)\nonumber\\
&+&\frac{2\eta}{8\pi\, r_{h1}\,l^{2}}\left(l^{2}(n-3)+(n-1)\,r_{h1}^{2}\right)\left(1-\frac{\omega\, r_{h1}^{n-2}}{2}\right).
\end{eqnarray}
The last two lines originate from the approximation $\exp(-S_{0})\approx 1-S_{0}=1-\omega\, r_{h}^{n-2}/2$, which is appropriate when the BH is small enough for the non-perturbative correction to matter. For large BHs, this approximation breaks down, but the exponential terms are already negligible at those scales.

The numerical evaluation of $\Delta F$ and $e^{-\Delta F}$ reveals a dimension-dependent behavior that is qualitatively different from the uncorrected case. In Table~\ref{tab:DeltaF}, we collect these quantities for various $r_{h2}$ with fixed $r_{h1}=1$, $\eta=1$, and $l=1$. For $n=3$, the free energy difference remains positive for all $r_{h2}<r_{h1}$, so that $e^{-\Delta F}<1$ throughout. For $n\geq 4$, however, $\Delta F$ turns negative at small $r_{h2}$, producing $e^{-\Delta F}>1$. The effect is dramatic in higher dimensions: for $n=10$ and $r_{h2}=0.1$, the free energy difference reaches $\Delta F\approx -4.31$, giving $e^{-\Delta F}\approx 74.5$. This means the Jarzynski weight $\langle e^{-\beta W}\rangle=e^{\beta\Delta F}$ is exponentially enhanced for evaporation to very small BHs in high dimensions, a direct consequence of the non-perturbative correction driving $G<0$ at small $r_{h}$ (cf.\ Table~\ref{tab:HP}).

\begin{table}[h!]
\centering
\renewcommand{\arraystretch}{1.5}
\caption{Free energy difference $\Delta F=F(r_{h2})-F(r_{h1})$ and $e^{-\Delta F}$ for the quantum-corrected ST--AdS BH, with $r_{h1}=1$, $\eta=1$, and $l=1$.}\label{tab:DeltaF}
\begin{tabular}{c c c c c c c c c}
\hline\hline
& \multicolumn{2}{c}{$n=3$} & \multicolumn{2}{c}{$n=4$} & \multicolumn{2}{c}{$n=5$} & \multicolumn{2}{c}{$n=10$}\\
\cmidrule(lr){2-3}\cmidrule(lr){4-5}\cmidrule(lr){6-7}\cmidrule(lr){8-9}
$r_{h2}$ & $\Delta F$ & $e^{-\Delta F}$ & $\Delta F$ & $e^{-\Delta F}$ & $\Delta F$ & $e^{-\Delta F}$ & $\Delta F$ & $e^{-\Delta F}$ \\
\hline
$0.10$ & $0.282$ & $0.754$ & $-0.381$ & $1.464$ & $-1.029$ & $2.797$ & $-4.310$ & $74.46$ \\
$0.20$ & $0.265$ & $0.767$ & $-0.004$ & $1.004$ & $-0.264$ & $1.302$ & $-1.597$ & $4.936$ \\
$0.30$ & $0.245$ & $0.783$ & $0.112$  & $0.894$ & $-0.028$ & $1.028$ & $-0.740$ & $2.096$ \\
$0.50$ & $0.196$ & $0.822$ & $0.188$  & $0.829$ & $0.142$  & $0.867$ & $-0.140$ & $1.151$ \\
$0.80$ & $0.092$ & $0.912$ & $0.144$  & $0.866$ & $0.181$  & $0.834$ & $0.089$  & $0.915$ \\
$0.90$ & $0.048$ & $0.953$ & $0.086$  & $0.918$ & $0.121$  & $0.886$ & $0.107$  & $0.898$ \\
$0.95$ & $0.025$ & $0.976$ & $0.046$  & $0.955$ & $0.069$  & $0.933$ & $0.080$  & $0.923$ \\
$1.00$ & $0$     & $1$     & $0$      & $1$     & $0$      & $1$     & $0$      & $1$     \\
\hline\hline
\end{tabular}
\end{table}

The physical significance of the sign reversal can be understood through the Jarzynski relation. When $\Delta F<0$, the final (smaller) BH state has lower free energy than the initial state, and the partition function ratio $Z_{2}/Z_{1}=e^{\beta\Delta F}>1$; the quantum work distribution is then biased toward positive work extracted from the gravitational field during evaporation. This phenomenon has no classical counterpart --- in the uncorrected thermodynamics ($\eta=0$), $\Delta F>0$ for all $r_{h2}<r_{h1}$ when $n\geq 4$, and no sign reversal occurs.

In Table~\ref{tab:Wavg}, we display the average work distribution $\langle W\rangle\approx -\Delta F$ (Jensen bound) for both $\eta=0$ and $\eta=1$. The quantum correction $\Delta\langle W\rangle=\langle W\rangle_{\eta=1}-\langle W\rangle_{\eta=0}$ is negligible for large $r_{h2}$ but grows rapidly as $r_{h2}$ decreases. The most striking feature is the sign change: for $n=4$ and $r_{h2}=0.1$, the uncorrected work distribution is $\langle W\rangle\approx -0.050$ (work done \emph{on} the BH), while the corrected value is $\langle W\rangle\approx +0.381$ (work \emph{extracted}), a reversal entirely due to the non-perturbative correction. For $n=10$ and $r_{h2}=0.1$, the corrected quantum work distribution reaches $\langle W\rangle\approx +4.31$, while the uncorrected value is essentially zero. This demonstrates that the quantum gravitational correction fundamentally alters the energetics of BH evaporation at the Planck scale, and the magnitude of the effect increases with the spacetime dimension.

\begin{table}[h!]
\centering
\renewcommand{\arraystretch}{1.5}
\caption{Average quantum work distribution $\langle W\rangle\approx -\Delta F$ (Jensen bound) for the ST--AdS BH, with $r_{h1}=1$ and $l=1$. The correction $\Delta\langle W\rangle=\langle W\rangle_{\eta=1}-\langle W\rangle_{\eta=0}$ quantifies the non-perturbative shift.}\label{tab:Wavg}
\begin{tabular}{c c c c c c c c c c}
\hline\hline
& \multicolumn{3}{c}{$n=3$} & \multicolumn{3}{c}{$n=4$} & \multicolumn{3}{c}{$n=5$} \\
\cmidrule(lr){2-4}\cmidrule(lr){5-7}\cmidrule(lr){8-10}
$r_{h2}$ & $\langle W\rangle_{\eta=0}$ & $\langle W\rangle_{\eta=1}$ & $\Delta\langle W\rangle$ & $\langle W\rangle_{\eta=0}$ & $\langle W\rangle_{\eta=1}$ & $\Delta\langle W\rangle$ & $\langle W\rangle_{\eta=0}$ & $\langle W\rangle_{\eta=1}$ & $\Delta\langle W\rangle$\\
\hline
$0.10$ & $-0.248$ & $-0.282$ & $-0.035$ & $-0.050$ & $\mathbf{+0.381}$ & $+0.431$ & $-0.008$ & $\mathbf{+1.029}$ & $+1.036$ \\
$0.30$ & $-0.228$ & $-0.245$ & $-0.018$ & $-0.137$ & $-0.112$ & $+0.025$ & $-0.064$ & $\mathbf{+0.028}$ & $+0.092$ \\
$0.50$ & $-0.188$ & $-0.196$ & $-0.008$ & $-0.188$ & $-0.188$ & $-0.000$ & $-0.147$ & $-0.142$ & $+0.005$ \\
$0.80$ & $-0.090$ & $-0.092$ & $-0.002$ & $-0.144$ & $-0.144$ & $-0.000$ & $-0.181$ & $-0.181$ & $-0.000$ \\
\hline\hline
\end{tabular}
\end{table}

In Fig.~\ref{fig-exp-delta}, we plot $e^{-\Delta F}$ as a function of $r_{h2}$ for fixed $r_{h1}=1$, $\eta=1$, and $l=1$. The exponential sensitivity of $e^{-\Delta F}$ to the free energy difference amplifies the effect of the quantum correction: even a modest shift in $\Delta F$ at small $r_{h2}$ produces a large change in $e^{-\Delta F}$, particularly in higher dimensions. For large $r_{h2}$ (close to $r_{h1}$), $\Delta F\to 0$ and $e^{-\Delta F}\to 1$, as expected. We can use the Jarzynski equality to relate the quantum work distribution to the partition functions as $\langle e^{-\beta W}\rangle=Z_{2}/Z_{1}$~\cite{6b}, so the relative weights of the partition functions depend on the quantum work done between the two states. The quantum gravitational corrections modify the equilibrium free energies entering the Jarzynski equality, which in turn changes the quantum work distribution. The data confirm that this modification is negligible for $r_{h2}\gtrsim 0.5$ but becomes dominant for $r_{h2}\lesssim 0.3$, with the crossover shifting to larger $r_{h2}$ in higher dimensions.

\begin{figure}[h!]
\begin{center}
\includegraphics[width=120mm]{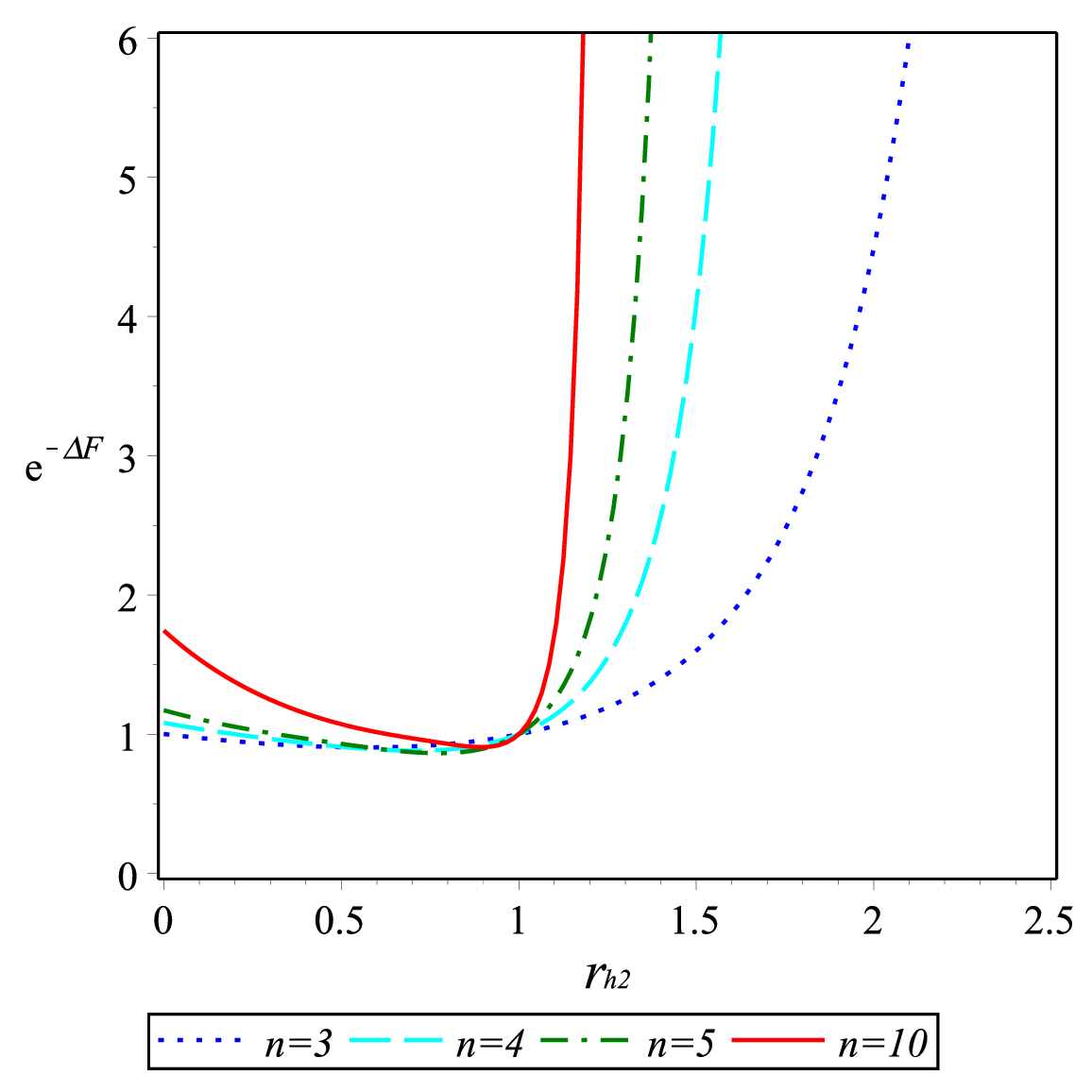}
\end{center}
\caption{$e^{-\Delta F}$ for the $n$-dimensional ST--AdS BH, with $r_{h1}=1$, $\eta=1$, and $l=1$.}
\label{fig-exp-delta}
\end{figure}


\section{Discussion and concluding remarks}\label{isec5}

In this paper, we have studied the non-perturbative quantum gravitational corrections to the thermodynamics and quantum work distribution of the $n$-dimensional ST--AdS BH. The corrected entropy $S=S_{0}+\eta\,e^{-S_{0}}$, with $S_{0}=\omega\, r_{h}^{n-2}/2$, modifies the equilibrium thermodynamic quantities through an exponential factor that is negligible for macroscopic BHs but becomes dominant when $S_{0}\lesssim 1$.

The corrected specific heat retains the classical divergence at $r_{h}^{*}=l\sqrt{(n-3)/(n-1)}$, which signals a second-order phase transition between stable large BHs and unstable small BHs for $n\geq 4$. The quantum correction enters as the multiplicative factor $(1-\eta\,e^{-S_{0}})$, which suppresses $C_{V}$ at small $r_{h}$: for $n=4$ and $l=1$, the specific heat at $r_{h}=0.2$ is reduced from $C_{V}=-0.64$ ($\eta=0$) to $C_{V}=-0.14$ ($\eta=1$), a reduction of nearly $78\%$. For $n=3$, the divergence is absent and $C_{V}$ remains positive for all $r_{h}$, so the three-dimensional AdS BH is stable regardless of the quantum correction.

In the extended phase space, we confirmed that the uncharged ST--AdS BH has no vdW-type critical point: the equation of state is a special case of the vdW equation with excluded-volume parameter $b=0$ and interaction parameter $a=(n-3)/[(n-2)\pi]$, and the isotherms are monotonically decreasing. The virial expansion terminates at the second coefficient $B(T)=-(n-3)/[(n-2)\pi T]$, with the Boyle temperature formally infinite. The only phase transition is of Hawking--Page type.

The analysis of the Gibbs free energy revealed a result that, to our knowledge, has not been reported previously. For $n\geq 4$ and $\eta=0$, the Gibbs free energy $G$ remains positive for all $r_{h}$, and no Hawking--Page transition takes place. When the non-perturbative correction is switched on ($\eta=1$), the exponential and incomplete gamma function terms drive $G$ negative at small $r_{h}$, creating a zero-crossing that corresponds to a quantum-induced Hawking--Page transition. The transition temperature increases with the spacetime dimension: $T_{\rm HP}\approx 0.446$ for $n=4$, $T_{\rm HP}\approx 0.610$ for $n=5$, and $T_{\rm HP}\approx 1.371$ for $n=10$ (all with $l=1$). This suggests that non-perturbative quantum corrections open a new thermodynamic channel that is absent in the semi-classical limit.

The corrected internal energy was obtained in closed form in terms of the upper incomplete gamma function $\Gamma(z,x)$, expressed through the Laguerre polynomial expansion. We noted and corrected a typographical error present in the original derivation, where $L_{0}^{(z)}(x)$ was written as $0$ instead of the standard value $1$. For $n=4$ and $r_{h}=0.1$, the quantum correction raises the internal energy from $E=0.101$ to $E=0.440$, an increase exceeding a factor of four.

The quantum work distribution during BH evaporation was derived using the Jarzynski equality, with the Jensen bound $\langle W\rangle\geq -\Delta F$ providing the leading estimate. The free energy difference $\Delta F$ between two BH states (initial $r_{h1}=1$, final $r_{h2}$) exhibits a sign reversal at small $r_{h2}$ for $n\geq 4$ when $\eta=1$: the corrected $\Delta F$ becomes negative, meaning the smaller BH has lower free energy. This sign reversal has no counterpart in the uncorrected thermodynamics and it fundamentally changes the energetics of evaporation. For $n=4$ and $r_{h2}=0.1$, the average quantum work distribution flips from $\langle W\rangle\approx -0.050$ (work done on the BH) to $\langle W\rangle\approx +0.381$ (work extracted). The effect grows with dimension: for $n=10$ and $r_{h2}=0.1$, the corrected quantum work distribution reaches $\langle W\rangle\approx +4.31$, while the uncorrected value is essentially zero. The exponential amplification of $e^{-\Delta F}$ (reaching $\approx 74.5$ for $n=10$ at $r_{h2}=0.1$) demonstrates that the partition function ratio $Z_{2}/Z_{1}$ is strongly modified by the quantum correction, with implications for the statistical weight of final-state configurations accessible through evaporation.

These results complement earlier studies of non-perturbative corrections to M2--M5 brane systems~\cite{6b1} and Myers--Perry BHs~\cite{6b2}, where similar exponential entropy corrections were found to modify the quantum work distribution. The new feature here is the dimension-dependent sign reversal of $\Delta F$ and $\langle W\rangle$, which arises because the ST--AdS geometry allows the correction to compete with the classical terms at small $r_{h}$ in a way that depends sensitively on $n$.

Our future plans about this study is as follows: First, the same non-perturbative correction applied to a charged (Reissner--Nordstr\"{o}m--AdS) BH would interact with the vdW-type phase transition and could modify the critical exponents near the critical point. Second, the formalism can be extended to rotating BHs in higher dimensions, where the interplay between the angular momentum and the exponential correction may produce new features in the phase diagram. Third, in the context of the AdS/CFT correspondence, the quantum-corrected partition function ratio $Z_{2}/Z_{1}$ has a dual interpretation in terms of the boundary CFT; it would be valuable to identify the CFT observable that captures the sign reversal of $\Delta F$. Finally, the non-perturbative correction can be incorporated into the quantum Raychaudhuri equation to study its effect on the focusing of geodesic congruences near the singularity, potentially modifying the conditions under which singularity theorems apply.

\section*{Acknowledgments}

\.{I}.~S. extends appreciation to T\"{U}B\.{I}TAK, ANKOS, and SCOAP3 for their financial assistance. Additionally, he acknowledges the support from COST Actions CA22113, CA21106, CA23130, CA21136, and CA23115, which have been pivotal in enhancing networking efforts. Fabiano F. Santos is partially supported by Conselho Nacional de Desenvolvimento Cient\'{\i}fico e Tecnol\'{o}gico (CNPq) under grant 302835/2024-5.

\section*{Data Availability Statement}

In this study, no new data was generated or analyzed.

}

\end{document}